\documentclass[fleqn, usenatbib]{mnras}

\usepackage[T1]{fontenc}

\DeclareRobustCommand{\VAN}[3]{#2}
\let\VANthebibliography\thebibliography
\def\thebibliography{\DeclareRobustCommand{\VAN}[3]{##3}\VANthebibliography}

\usepackage{graphicx}	
\usepackage{amsmath}	
\usepackage{amssymb}	
\usepackage{xcolor}
\usepackage{newtxtext,newtxmath}

\title[BCG Evolution]{Brightest Cluster Galaxies Are Statistically Special From $z=0.3$ to $z=1$}

\author[R. Dalal et al.]{
Roohi Dalal$^{1}$\thanks{E-mail: rdalal@princeton.edu},
Michael A. Strauss$^{1}$,
Tomomi Sunayama$^{2}$,
Masamune Oguri$^{3,4,5}$,
Yen-Ting Lin$^{6}$,
\newauthor Song Huang$^{1}$,
Youngsoo Park$^{5}$
and Masahiro Takada$^{5}$
\\
$^{1}$Department of Astrophysical Sciences, Princeton University, Peyton Hall, Princeton, NJ 08544, USA\\
$^{2}$Division of Particle and Astrophysical Science, Graduate School of Science, Nagoya University, Nagoya 464-8602, Japan\\
$^{3}$Research Center for the Early Universe, University of Tokyo, Tokyo, 113-0033, Japan\\
$^{4}$Department of Physics, University of Tokyo, Tokyo 113-0033, Japan\\
$^{5}$Kavli Institute for the Physics and Mathematics of the Universe
(Kavli IPMU, WPI), University of Tokyo, Chiba 277-8582, Japan\\
$^{6}$Institute of Astronomy and Astrophysics, Academia Sinica, Taipei 10617, Taiwan\\
}

\date{Accepted XXX. Received YYY; in original form ZZZ}

\pubyear{2021}

\begin{document}
\label{firstpage}
\pagerange{\pageref{firstpage}--\pageref{lastpage}}
\maketitle

\begin{abstract}
We study Brightest Cluster Galaxies (BCGs) in $\sim5000$ galaxy clusters from the Hyper Suprime-Cam (HSC) Subaru Strategic Program. The sample is selected over an area of 830 $\textrm{deg}^2$ and is uniformly distributed in redshift over the range $z=0.3-1.0$. The clusters have stellar masses in the range $10^{11.8} - 10^{12.9} M_{\odot}$. We compare the stellar mass of the BCGs in each cluster to what we would expect if their masses were drawn from the mass distribution of the other member galaxies of the clusters. The BCGs are found to be "special", in the sense that they are not consistent with being a statistical extreme of the mass distribution of other cluster galaxies. This result is robust over the full range of cluster stellar masses and redshifts in the sample, indicating that BCGs are special up to a redshift of $z=1.0$. However, BCGs with a large separation from the center of the cluster are found to be consistent with being statistical extremes of the cluster member mass distribution. We discuss the implications of these findings for BCG formation scenarios.
\end{abstract}




\section{Introduction}

Brightest Cluster Galaxies (BCGs), the most luminous galaxies in clusters, are typically old and red elliptical galaxies which lie near the center of the cluster. These galaxies appear to be special or distinct from other cluster galaxies in many ways. They are generally uniform in color \citep{schneider1983a}, have a narrow distribution of luminosities \citep{schneider1983b, sandage1972b, postman_lauer1995}, sit at the bottom of the cluster potential well \citep{quintana1982, jones1984, lin2004}, are aligned with their parent cluster \citep{sastry1968, binggeli1982, lambas1988, niederste-ostholt2010, hao2011}, and have extended envelopes \citep{gonzalez2005, zibetti2005, huang2018}, a higher [$\alpha$/Fe] than their satellite galaxies \citep{gu2018}, as well as small peculiar velocities relative to the cluster mean \citep{postman_lauer1995, vandenBosch2005, coziol2009}. Many of these properties of BCGs allow for their use in a number of cosmological contexts. The high and nearly standard luminosities of BCGs have been used to study the Hubble flow out to large distances \citep{sandage1972a, sandage1972b, gunn1975}. However, \citet{tinsley1976} showed that uncertainties in stellar population evolution limit the use of BCGs as standard candles. Their potential as standard candles was improved when \citet{hoessel1980} showed that the BCG metric luminosity, $L_m$, is correlated with the logarithmic slope of the curve of growth, $\alpha$. The $L_m - \alpha$ relation avoids the complications of characterizing and measuring the light from the extended envelopes of BCGs, enabling their use for more precise measurements of the Hubble flow and the local velocity field \citep{lauer1994, lauer2014, rozo2010}.

One explanation for the special properties of BCGs is the process of cannibalism, whereby a BCG in the center of the cluster potential tends to engulf and merge with its neighbors, causing it to have a very high luminosity \citep{ostriker1975, hausman1978}. Because dynamical friction is a mass-dependent process, massive galaxies will tend to merge quickly with the central galaxy. Such major mergers would then build up the mass of the central galaxy while also reducing the number of massive satellite galaxies. \citet{lauer1988} estimated a cannibalism rate for 8 Abell cluster BCGs of $\sim 2L_*$ per $5 \times 10^9$ yr by looking for evidence of interactions between the BCG and secondary galaxies. However, \citet{merritt1985} claimed that cannibalism is not a viable mechanism for BCG evolution, since tidal stripping would reduce the masses of the cluster galaxies, causing dynamical friction to be too slow for the BCG to capture other galaxies. BCGs might also grow via cooling flows, i.e. the radiative cooling of the intra-cluster medium, causing it to sink to the bottom of the cluster potential well, where the BCG is often located \citep{cowie1977, fabian1977}. We want to understand whether BCGs evolved to become special as a result of their environment (via cannibalism or cooling flows, for example), or whether they were born with their special properties. If the former is the case, we would like to determine the epoch at which the processes that made BCGs special took place.

Theoretical predictions for BCG evolution over cosmic time have typically utilized semi-analytical models of galaxy formation, as full cosmological hydrodynamical simulations would be very computationally expensive over the large volumes needed to study rare objects like BCGs. Semi-analytical models allow one to estimate the mass growth of BCGs over time (for example, see \citet{Aragon-Salamanca1998} and \citet{DeLucia2007}. However, these models, by construction, have shock heated gas cooling radiatively and condensing only onto the central galaxy of a halo, i.e. the BCG. Furthermore, these models usually only account for mergers between the satellite galaxies and the central galaxy, neglecting mergers between different satellite galaxies. Thus, such semi-analytical models make BCGs special by design. Alternative methods of modeling BCG evolution, such as \citet{dubinksi1998}, \citet{laporte2015} and \citet{ruszkowski2009}, use pure $N$-body simulations and only populate the dark matter halos with stellar subsystems below a redshift of $z \sim 2$. However, such methods assume that only dry mergers (i.e. those that are gas poor and with little star formation) are responsible for mass assembly in the inner parts of clusters. More recently, \citet{Ragone-Figueroa2018} used cosmological hydrodynamical simulations to study BCG growth in massive galaxy clusters and found that at high redshifts ($z\gtrsim 2$), global star formation in massive satellite galaxies is more important than in-situ star formation in the main progenitor. At lower redshifts, in-situ star formation becomes increasingly important, with the assembly of about half of the BCG mass happening at redshifts around $\sim 1.5$. They also found a decrease in assembly redshifts and increase in mass growth factors with aperture, suggesting that an \textit{inside-out} scenario for BCG growth is likely. We aim to use observational data to contribute to our understanding of BCG evolution in terms of the development with time of their special nature.

One way to characterize the special nature of BCGs, motivated by the narrow distribution of their luminosities, is by determining whether these luminosities are consistent with being a statistical extreme of the cluster member luminosity distribution. If their luminosities are higher than one would expect from the luminosity distribution of other galaxies, then they must follow their own special distribution. Early studies found that the BCG luminosities are consistent with simply being statistical extremes \citep{peebles1968, geller1976, geller1983}. Other studies have concluded that BCGs comprise an entirely different population from other cluster members. \citet{schneider1983b} showed that the BCGs in the cores of 83 Abell clusters have a small spread in absolute magnitude, $\sigma(m_1) \approx 0.35$, compared to that of the second- and third-brightest galaxies in their cluster sample, $\sigma(m_2) \approx 0.55$ and $\sigma(m_3) \approx 0.65$. If BCGs were drawn from the same luminosity distribution as other galaxies in the cluster, we would expect them to have a larger spread in magnitude, similar to the second- and third-ranked galaxy. The small spread in BCG magnitudes was corroborated by \citet{postman_lauer1995}, who found a scatter in the BCG metric luminosity of 0.327 mag. \citet{bhavsar1985} used BCG photometry from \citet{hoessel1980} to compare the distribution of BCG luminosities to Fisher-Tippet extreme value distributions \citep{fisher1928}. They concluded that BCGs are not simply a statistical extreme of the luminosity distribution of other cluster galaxies, and thus must be special. \citet{tremaine_richstone1977} proposed a related test of BCG specialness which involves the difference in luminosities between the BCG and second-brightest galaxy in the cluster. One would expect this difference to be small if both galaxies are drawn from the same luminosity distribution. \citet{loh_strauss2006} applied this test to a sample of luminous red galaxies from 2099 $\textrm{deg}^2$ of SDSS imaging data and found that the difference in luminosities is large enough to infer that BCGs are special. This conclusion was supported by \citet{shen2014} who used order statistics to conclude that BCGs in the SDSS DR7 catalog are special, as they are $\sim 0.2$ mag brighter on average than the expected luminosity from order statistics. \citet{lin_etal2010} applied the \citet{tremaine_richstone1977} test to clusters from the C4 spectroscopic catalog \citep{miller2005}, but found that the two statistics proposed by \citet{tremaine_richstone1977} gave conflicting results. \citet{lin_etal2010} went on to use the $M_{\textrm{bcg}}-M_{\textrm{tot}}$ correlation for the observed clusters, compared to that for a simulated cluster sample, to conclude that BCGs in high luminosity clusters are special, while those in less luminous clusters are consistent with being a statistical extreme of the underlying population, a notable difference from the results of previous studies. On the other hand, \citet{dobos2011} applied order statistics to the bright end of the luminosity distribution of spectroscopically identified early-type galaxies in the SDSS DR7 catalog, and concluded that BCGs can be considered statistical extremes when galaxies are binned by redshift, rather than considering each cluster separately. \citet{paranjape2012} also used order statistics with a galaxy group catalog from \citet{berlind2006} to conclude that BCG luminosities are statistical extremes of the group galaxy luminosity function. The question of whether BCGs masses or luminosities are indeed special remains unsettled, particularly as a function of redshift and cluster mass. 

In this paper, we study the redshift evolution of the statistical nature of BCGs using a cluster catalog from the Hyper Suprime-Cam (HSC) survey \citep{aihara2018a}, constructed using the CAMIRA algorithm \citep[Cluster finding Algorithm based on Multiband Identification of Red-sequence gAlaxies;][]{oguri2014, oguri_2017}. The depth of the survey, combined with its large solid angle and excellent photometry, allows us to uniformly select a large cluster sample out to large redshifts. This enables us to test the statistical nature of BCGs as a function of redshift. 

Although the studies discussed above focus on BCG luminosities relative to a cluster galaxy luminosity distribution, we discuss the statistical nature of BCGs in terms of their stellar masses and the distribution of masses of other cluster galaxies. Although the stellar mass is an inferred quantity, rather than an observable, it is a more fundamental quantity, uncomplicated by issues of star formation histories and stellar ages. Furthermore, the stellar mass of central galaxies has been shown to correlate better with host properties than stellar luminosity \citep{more2011}. 

The definition of the BCG can vary from study to study, with some considering proximity to the center of the cluster potential to be a defining characteristic, and therefore choosing the BCG to be the most luminous object near the center of the cluster. However, \citet{skibba2011} find that the BCG is often not the central galaxy, with the fraction of non-central BCGs varying from $\sim 0.25$ to $\sim 0.4$, depending on the halo mass. We want to explore the special nature of BCGs in terms of their masses, and therefore use the term Brightest Cluster Galaxy quite literally, defining our BCGs to be the most luminous, or equivalently, the most massive galaxies in each cluster, regardless of position. 

We apply two different methods to quantify the deviation of the BCG mass from what would be expected given the cluster member mass distribution. First, we use the magnitude-gap statistics proposed by \citet{tremaine_richstone1977}. We also apply the method proposed by \citet{lin_etal2010}, which compares the observed BCG mass distribution to mock realizations of the mass distribution which we would expect if BCGs were simply statistical extremes of the cluster member mass distribution. These two approaches are complementary, as the former assumes a power law shape for the bright-end of the luminosity (or mass) function, but nothing about the universality of the distribution across clusters. The latter assumes that the luminosity function is universal, but makes no assumptions about its shape.

The structure of this paper is as follows. In Section~\ref{sec:sample}, we describe our cluster sample and the galaxy data. We describe the two tests of the statistical nature of BCGs in Section~\ref{sec:methods} and the results from each in Section~\ref{sec:results}. In Section~\ref{sec:systematics} we discuss tests of the robustness of our results. We conclude by discussing the implications of our findings on BCG formation scenarios in Section~\ref{sec:discussion}.

Throughout this paper, we adopt a flat $\Lambda$CDM cosmological model where $\Omega_m = 0.3$, $\Omega_{\Lambda} = 0.7$ and $H_0 =70 \ \textrm{km} \ \textrm{s}^{-1} \textrm{Mpc}^{-1}$.

\section{Cluster Sample}
\label{sec:sample}

\subsection{The HSC-SSP Survey}

The Hyper Suprime-Cam Subaru Strategic Program (HSC-SSP survey) is an ongoing imaging survey using the 8.2 m Subaru Telescope \citep{aihara2018a}. The Hyper Suprime-Cam is a wide-field camera with 870 Megapixels covering a 1.5 deg diameter field of view \citep{miyazaki2012, miyazaki2018}. The survey consists of a wide, a deep and an ultradeep layer, each observed in the \textit{grizy} broadband filters \citep{kawanomoto2018}. The median seeing for the wide layer in the \textit{i} band is 0.6 arcsec, allowing for excellent image quality. In this paper, we use the wide survey, which aims to cover 1400 $\textrm{deg}^2$ with a point source $5\sigma$ depth of $r \sim 26$ mag. Our cluster sample is based on the internal S20A data release, which covers an area of 830 $\textrm{deg}^2$.

The survey data is reduced by a pipeline, presented in \citet{bosch_2018}, that has been developed in parallel with the pipeline for the Vera C. Rubin Observatory \citep{ivezic2019, juric2017}. Astrometric and photometric calibrations are carried out by comparison with data from the PanSTARRS1 survey \citep{chambers2016}. There have been two public data releases from the survey, presented in \citet{aihara2018b, aihara2019}. A notable difference between the pipeline version used for the S20A data and that used for Public Data Release 2 is that the former no longer uses the global sky subtraction algorithm (described in Section 4.1 of \citealt{aihara2019}) for object detection, deblending and measurements. This results in better behavior of the deblender around bright objects and in dense environments, so we expect improved accuracy of the photometry in clusters in the S20A data. 

\subsection{Cluster and Galaxy Data}
\label{sec:cluster_gal_data}

The galaxy clusters in this work are identified by applying the CAMIRA algorithm to object catalogs from the S20A internal data release of the HSC survey. We summarize the algorithm here, and refer the reader to \citet{oguri2014} and \citet{oguri_2017} for a more detailed description. 

CAMIRA is a matched-filter cluster finding algorithm. It calibrates a stellar population synthesis (SPS) model using a sample of spectroscopically observed galaxies from a number of surveys, and then uses this model to search for overdensities of red-sequence galaxies at an assumed cluster location and redshift. The algorithm uses the SPS model of \citet{bruzual2003}, which characterizes the properties of galaxies by a number of parameters, including the age of the galaxy, its star formation history, its metallicity, its stellar mass and its dust extinction. The algorithm adjusts these parameters to reproduce the observed colors of red sequence galaxies, and uses the model to calculate the likelihood of galaxies being on the red sequence as a function of redshift. The model also gives a mass-to-light ratio for each galaxy. Since the CAMIRA algorithm looks for red-sequence galaxies, i.e. galaxies with old stellar populations, we expect the galaxies in our sample to have similar mass-to-light ratios, and so we expect the mass distribution of these galaxies to have a similar shape to the luminosity distribution. This allows us to employ tests of the special nature of BCGs that have historically been based on luminosity using the masses instead. 

Each galaxy is fit separately to an exponential and a de Vaucouleurs profile. The image is then fit to a linear combination of the two, giving the \texttt{cmodel} magnitude \citep{abazajian2004, bosch_2018} reported by the HSC pipeline. Together, the mass-to-light ratio from the SPS model derived assuming the \citet{salpeter1955} initial mass function and the \texttt{cmodel} magnitude measured by the HSC pipeline give a stellar mass for each galaxy. We note that the \texttt{cmodel} magnitude likely does not capture the total magnitude of the galaxy leading to an underestimation of its stellar mass, particularly in the case of a galaxy like a BCG with an extended envelope. We show in Section~\ref{sec:blendedness}, however, that this does not lead to biases in the statistics employed here. The algorithm then computes a three-dimensional richness map, based on Equation 11 of \citet{oguri2014}, which it uses to identify cluster candidates. For each peak in the richness map, the candidate cluster is assigned a photometric redshift by maximizing a weighted likelihood of each galaxy being a red sequence galaxy at a redshift $z$ based on the SPS model. The cluster redshift estimate is refined by identifying a massive, high-significance galaxy near the peak of the richness map as the central galaxy, and then re-estimating the cluster redshift using the central galaxy position as the peak position. This process is repeated until the solution converges. The photometric redshifts inferred by this method are quite accurate, with a bias and scatter in $\Delta z/(1+z)$ of 0.005 and 0.01, respectively, over the redshift range of the sample.

After the final cluster candidate is obtained, each galaxy in the cluster is assigned a weight, $w$, representing its membership probability. This membership probability is based on the likelihood of it being a red sequence galaxy at redshift $z$, a stellar mass filter, which is $\sim 1$ for the high-mass galaxies that we are considering here, and a compensated spatial filter, which favors galaxies with a smaller projected distance from the cluster center. The spatial filter is close to unity for galaxies within $\sim 0.7$~Mpc of the cluster center. The cluster richness, $N_{\textrm{mem}}$ is defined to be the sum of these weights for all possible member galaxies, with an additional masking correction term to account for regions of the sky that are masked due to survey boundaries and bright stars. We impose a richness threshold of $N_{\textrm{mem}}>15$ for our sample. The cluster stellar mass is the sum of the masses of the member galaxies, weighted by their membership probability. Note that we do not measure the cluster halo mass \citep[see][for the correlation between the richness and halo mass for CAMIRA clusters]{murata2019,chiu2020a,chiu2020b}. 

\begin{figure}
    \centering
    \includegraphics[width=\columnwidth]{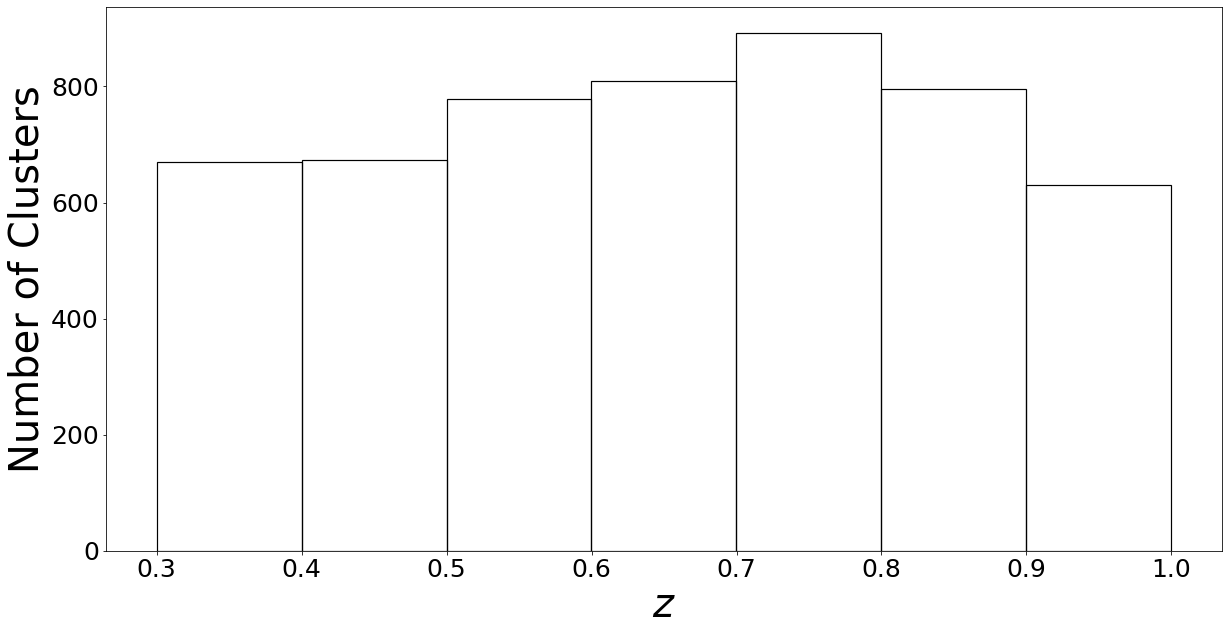}
    \caption{Redshift histogram of the CAMIRA sample. The distribution of clusters is close to flat over the redshift range we consider in this study, $0.3 \leq z<1.0$.}
    \label{fig:cluster_z_dist}
\end{figure}

This results in a sample of 5731 clusters with $N_{\textrm{mem}}>15$ with redshifts in the range $0.3 \leq z < 1.0$ from the internal S20A data release. The clusters have stellar masses in the range $10^{11.8} - 10^{12.9} M_{\odot}$ and richnesses in the range $15<N_{\textrm{mem}}\leq110$, corresponding to a halo mass of approximately $10^{14.4} - 10^{15.3} M_{\odot}$, based on the mass-richness relation found by \citet{murata2018}. The depth of the HSC survey, which has a limiting magnitude of $\sim 26$~mag, means that the galaxy sample probes the same range of luminosities ($L \gtrsim 0.2L_*$), and hence galaxy masses ($M \gtrsim 10^{10.2}M_{\odot})$, up to a redshift $z \sim 1$, so the distribution of clusters is flat over our redshift range. Figure~\ref{fig:cluster_z_dist} shows the number of clusters in the sample as a function of redshift, in redshift bins of width $z=0.1$. There are two versions of the CAMIRA catalog - one which uses a bright-star mask provided by the HSC database, and one which does not. We use the version without the star mask in the analysis below, but we have repeated these tests on the version with the star mask, and have found no significant differences. 

Since the CAMIRA algorithm is a red sequence cluster finder, there is a possibility that we are missing blue, star-forming BCGs. We do not expect the fraction of blue BCGs to be very high (for example, see \citealt{pipino2011}), although this fraction could increase at redshift closer to $z\sim1$ \citep{mcdonald2016}. 

In our exploration of the statistical nature of BCG masses, we do not use the central galaxy which is identified by CAMIRA, as this is based on both the mass of the galaxy and its proximity to the center of the cluster. Instead, we define BCGs to be the most massive galaxy with a membership probability $w>0.1$ in a given cluster. Due to these different definitions, the BCGs used in this study are not always at the center of their host clusters. The effect of this assumption is explored in Section \ref{sec:offset}. 

Since the masses of the galaxies are estimated using the \texttt{cmodel} magnitudes, we would expect these masses to have large uncertainties in highly blended scenarios, and to be susceptible to errors from the deblending process. The high-density nature of clusters, and particularly cluster centers, makes it likely that BCGs will be blended with other galaxies in the field. In order to understand whether we should be concerned about inaccuracies in BCG masses due to deblending details, we use the absolute blendedness parameter, $b_{\textrm{abs}}$, in each band for each of the identified BCGs \citep{bosch_2018}. $b_{\textrm{abs}}$ represents the fraction of light from a given object that is affected by blending with other sources. It is an upper limit for how incorrect the photometry of a galaxy might be due to blending. If the deblender works perfectly, then the photometry is unaffected regardless of the blending fraction. In each band, about 90\% of BCGs have $b_{\textrm{abs}}<0.1$, suggesting that we do not need to be too concerned about the effects of the deblending process. We remove clusters that have a BCG with \textit{i}-band $b_{\textrm{abs}} > 0.1$ from our analysis (about 463 clusters, or 8.1\% of our total sample). We explore the effect of removing this blendedness cut, as well as making it more stringent, in Section~\ref{sec:blendedness}. Additionally, we find 8 pairs of clusters that share the same BCG. Although this is allowed by the CAMIRA algorithm, we remove these clusters from our sample. This leaves us with a final sample of 5250 clusters. 

\section{Statistical Tests of BCG Dominance}
\label{sec:methods}

Historically, tests of the statistical nature of BCGs have focused on the luminosity of BCGs and their relation to the luminosity distribution of other galaxies in the cluster. As mentioned above, in this study we use stellar mass, a more fundamental quantity. However, we begin by describing previously used tests of BCG specialness in terms of luminosity, keeping in mind that the uniform mass-to-light ratio for the old, red galaxies in our sample means that we can use mass and luminosity interchangeably. Note that we use $m$ to represent absolute magnitude, and $M$ to represent stellar mass. 

The \citet{tremaine_richstone1977} statistics, described below, rely on two assumptions about the nature of the galaxy luminosity distribution in clusters. The first is that the luminosity function follows the model of \citet{scott1957}, i.e. for a given luminosity distribution, overlapping magnitude intervals are statistically independent. For a cumulative luminosity function $\psi(m)$, where $m$ is the absolute magnitude,

\begin{equation}
    \psi(m) = \int_{-\infty}^m \phi(m') \textrm{d}m',
\end{equation}
the probability of finding $\nu$ galaxies in the magnitude interval [$m_a$, $m_b$] is given by Poisson statistics,

\begin{equation}
    p_{\nu} (m_a, m_b) = \frac{\left[\psi(m_b) - \psi(m_a)\right]^{\nu}}{\nu!} \exp{[\psi(m_a) -\psi(m_b)]}.
\end{equation}
The probability of finding one galaxy in the infinitesimal interval [$m$, $m + \textrm{d}m$] is then

\begin{equation}
    \textrm{d}P = \phi(m)\textrm{d}m.
\end{equation}
So, the probability distribution of the $j$th-brightest galaxy having a magnitude $m$ is the probability of $j-1$ galaxies being brighter than $m$ and having one galaxy in the interval [$m$, $m + \textrm{d}m$]

\begin{equation}
\begin{aligned}
    p_{(j)} \textrm{d}m &= p_{j-1}(-\infty, m) \times p_1(m, m+\textrm{d}m) \\
    & = \frac{\psi(m)^{j-1}}{(j-1)!} \exp{\left[-\psi(m)\right]\phi(m) \textrm{d}m}
\end{aligned}
\end{equation}

The second assumption on which the Tremaine-Richstone statistics hinge is that the bright end of the luminosity function drops off as a power law in  luminosity, or equivalently, an exponential in magnitude. That is, we can assume a universal bright end differential luminosity function in the form

\begin{equation}
    \phi(m) \simeq \exp{[\alpha(m-m_0)]}
\end{equation}
where $\alpha$, which is negative, parameterizes the steepness of the function and $m_0$ gives the normalization of the luminosity function in any given cluster.

\subsection{Tremaine-Richstone Statistics}
\label{sec:t_stats_method}

\citet{tremaine_richstone1977} show that the difference in magnitudes between the first- and second-ranked member galaxies of a cluster can be used to determine whether both of these galaxies are drawn from the same power-law luminosity function ($\phi \sim \exp{(\alpha m)}$). The average magnitude difference between the first- and second-ranked galaxy, $m_{12} \equiv m_2-m_1$, cannot be too large if both galaxies are drawn from the same luminosity function. The specific statistics proposed by Tremaine \& Richstone are:
\begin{equation}
    T_1 \equiv \frac{\sigma(m_1)}{\left<m_{12}\right>} \geq 1.
\end{equation}
and 
\begin{equation}
    T_2 \equiv \frac{\sigma(m_{12})}{\left<m_{12}\right>} \gtrsim 0.82,
\end{equation}
where $\sigma(m_1)$ is the standard deviation of the BCG magnitudes over the observed cluster sample, $\left<m_{12}\right>$ is the mean of the difference in magnitudes between the first- and second-ranked galaxy in each cluster, and $\sigma(m_{12})$ is the standard deviation of this difference. The limits of 1 and 0.82 (above which BCGs are consistent with being statistical extremes) are derived from their assumptions about the nature of the galaxy luminosity distribution. These statistics indicate that for the two galaxies to be drawn from the same luminosity function, the mean magnitude difference must be of the same order as the spread of the first-ranked galaxy magnitudes, and the spread of the magnitude difference must also be of the same order as the mean difference. 

Although this test was designed with galaxy magnitudes in mind, the same statistics and inequalities can be used with stellar mass distributions instead of magnitude distributions. Assuming a constant mass-to-light ratio for our galaxy sample, the shape of the mass function on the bright end should be similar to that of the luminosity function. So, we can use these same statistics with the log of the stellar masses, rather than magnitudes.

\begin{equation}
\label{eq:t_1}
    T_1 \equiv \frac{\sigma(\log_{10}(M_1))}{\left<\log_{10}(M_1/M_2)\right>} \geq 1.
\end{equation}
and 
\begin{equation}
\label{eq:t_2}
    T_2 \equiv \frac{\sigma(\log_{10}(M_1/M_2))}{\left<\log_{10}(M_1/M_2)\right>} \gtrsim 0.82.
\end{equation}
where $M_1$ is the stellar mass of the first-ranked galaxy (the BCG) and $M_2$ is the stellar mass of the second-ranked galaxy.
 
We select the first- and second-ranked galaxies from the CAMIRA catalog list of member galaxies in each cluster by finding the most massive and second-most massive galaxy in each cluster with a minimum membership probability $w>0.1$. We explore the effect of changing this threshold in Section~\ref{sec:weights}.  

We compute the $T_1$ and $T_2$ statistics for our full cluster sample, as well as in redshift bins of width $\Delta z = 0.1$, to search for evidence of redshift evolution in the special nature of BCGs, as well as to ensure that we are not sensitive to redshift-dependent selection effects. 

The lower limits on $T_1$ and $T_2$ in Equations~\ref{eq:t_1} and \ref{eq:t_2} are derived by \citet{tremaine_richstone1977} by assuming that the luminosity distribution has a power law form. However, this is not necessarily the case for our data. For example, observations of galaxies, including \citet{blanton2003}, have shown that the optical luminosity function is better described by an exponential distribution at the high-luminosity end \citep{schechter1976}. We can repeat the Tremaine-Richstone test without assuming any particular form of the luminosity (or mass) distribution by using the galaxies in our data set to construct a comparison sample in which BCGs are assumed to not be special, and comparing the resulting $T_1$ and $T_2$ values to the ones obtained from our data set. Specifically, we populate mock realizations of our cluster sample by drawing galaxies from a pool consisting of all member galaxies in our sample, and thereby assuming that the BCG of any mock cluster has a mass drawn from the overall mass distribution of cluster member galaxies. This process is described in further detail in Section~\ref{sec:mass_corr_method}. This allows us to use the Tremaine-Richstone statistics to determine whether BCGs are special without assuming a form for the mass (or luminosity) distribution of cluster member galaxies.

\subsection{\texorpdfstring{$M_{\textrm{bcg}}-M_{\textrm{tot}}$} \ \ correlation}
\label{sec:mass_corr_method}

\citet{lin_etal2010} show that the two Tremaine-Richstone statistics do not always lead to the same conclusion about whether BCGs are special or statistical extremes. When applying these statistics to samples of low and high mass clusters from the C4 spectroscopic catalog \citep{miller2005} from the fifth data release of the SDSS \citep{adelman-mccarthy2007}, they find that based on $T_1$, the BCGs in low mass clusters would appear to be "special" i.e., they violate Equation~\ref{eq:t_1} and therefore do not follow the luminosity function of other galaxies. However, using $T_2$ would lead one to conclude that it is the BCGs in high mass systems that are the special ones. This suggests that the $T_i$ statistics, when considered together, may not unambiguously determine the statistical nature of BCGs with respect to cluster mass. 

For this reason, we also try using an alternative test of the BCG luminosity function, following the method employed by \citet{lin_etal2010}. This involves studying the BCG luminosity-cluster luminosity correlation, or in our case, the correlation between the BCG stellar mass, $M_{\textrm{bcg}}$, and the cluster stellar mass, $M_{\textrm{tot}}$. We create mock realizations of this correlation (assuming that BCGs are statistical extremes of the cluster member mass distribution) and compare it to the observed correlation. 

We first create an overall galaxy pool, containing all member galaxies in our cluster sample with a weight parameter $w>0.1$. We discuss the effect of changing the minimum weight in Section~\ref{sec:weights}. For each cluster in the sample, we create a corresponding mock cluster by randomly drawing galaxies from the overall galaxy pool and adding their weighted mass $w_i M_i$ to the cluster. This is done until the total stellar mass of the mock cluster matches the observed cluster mass $M_{\textrm{tot}}$, as tabulated in the CAMIRA catalog. If the first $N$ galaxies give a total stellar mass $M_N < M_{\textrm{tot}}$, then the ($N+1$)th galaxy to bring the mock cluster mass to $M_{N+1} > M_{\textrm{tot}}$ is included if $M_{N+1} - M_{\textrm{tot}} < M_{\textrm{tot}}-M_{N}$. A galaxy drawn from the pool is considered for inclusion in the cluster with a probability equal to its weight (a galaxy with $w=1$ will always be included, while a galaxy with a weight of 0.1 will be included 10\% of the time). 

We also impose the richness cut of $N_{\textrm{mem}} >15$ that we used to select CAMIRA clusters. Without this cut, we end up with a large number of low-mass simulated clusters with a richness that is significantly smaller than 15. If a simulated cluster's richness is not above this value, it is discarded and re-populated until it matches the mass of the corresponding CAMIRA cluster and falls above the richness cut. We then choose the galaxy with the largest unweighted mass to be the BCG of the simulated cluster. We then divide the simulated cluster sample into 15 equi-populated bins of cluster stellar mass, and find the median BCG mass in each bin, $\overline{M_{\textrm{bcg, sim}}}$. We then take the median of $\overline{M_{\textrm{bcg, sim}}}$ over 200 Monte Carlo realizations of the mock cluster ensemble, $\left<\overline{M_{\textrm{bcg, sim}}}\right>$. This gives us the simulated $M_{\textrm{bcg, sim}}-M_{\textrm{tot}}$ correlation which we can compare to the correlation from our actual data. As in the case of the Tremaine-Richstone statistics, we carry out this exercise for the full sample as well as in bins of redshift.

\section{Results}
\label{sec:results}

\subsection{Tremaine-Richstone Statistics}
\label{sec:t_stats_results}

\begin{figure}
    \centering
    \includegraphics[width=\columnwidth]{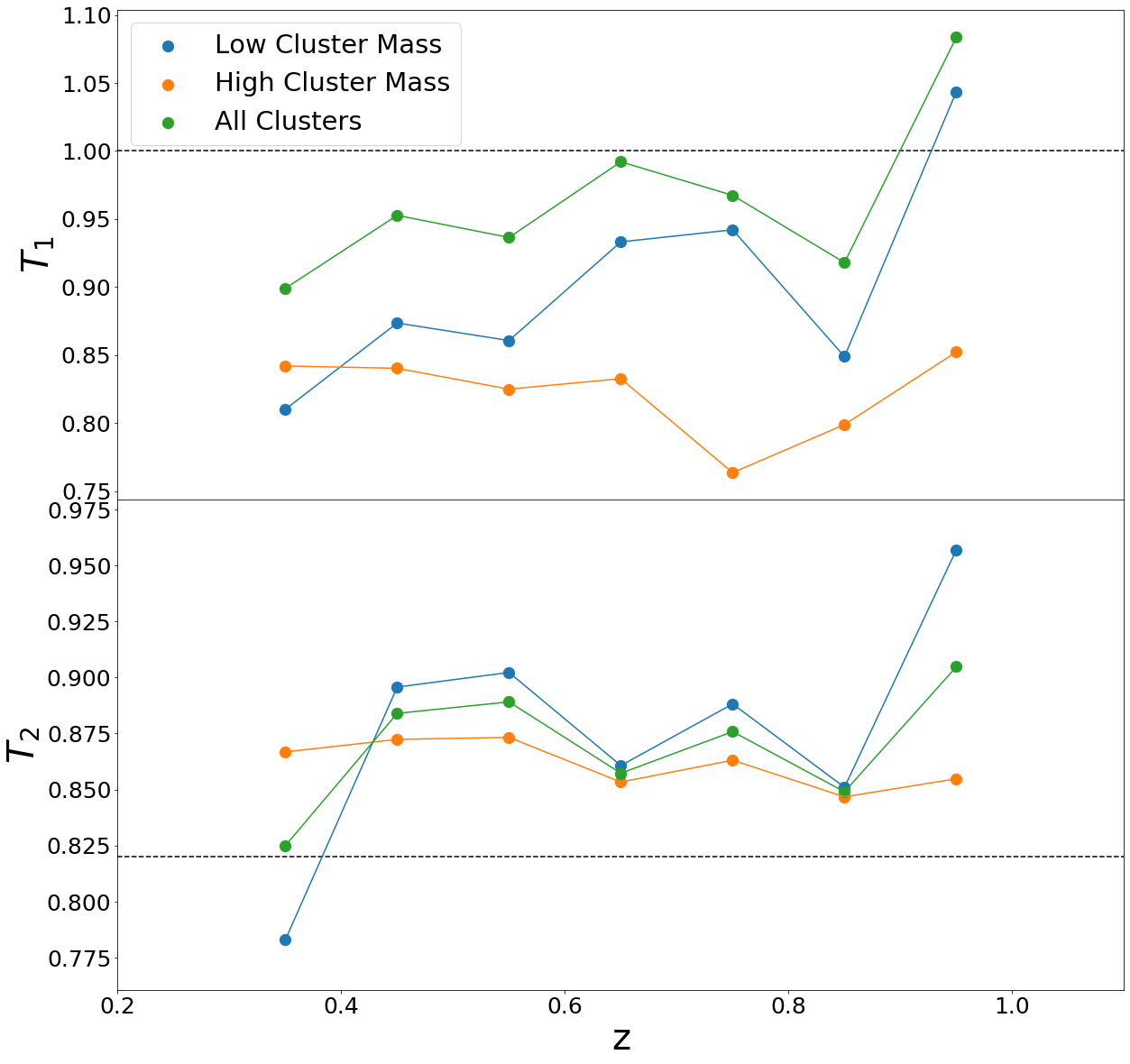}
    \caption{Values of the Tremaine-Richstone statistics for BCGs in redshift bins of width $\Delta z = 0.1$. The top panel shows the redshift evolution of $T_1$, while the bottom panel shows that of $T_2$. We also use the median cluster stellar mass in each bin (which does not vary much with redshift) to split the clusters in each bin into low mass and high mass subsamples. We plot the statistical limit as a dashed line, where values above that line would be consistent with BCGs being statistical extremes of the cluster member mass distribution. We see that based on the $T_1$ statistic, one would generally conclude that BCGs are `special', whereas the $T_2$ statistic indicates that BCGs are simply statistical extremes. The values of $T_1$ are lower for the cluster mass-based subsamples than for the full sample because the BCG mass is correlated with cluster stellar mass, which causes $\sigma(\log_{10}(M_1))$ to drop when splitting the clusters by stellar mass.}
    \label{fig:t_stats_z_evol}
\end{figure}

Using our full cluster sample, we find $\sigma(\log_{10}(M_1)) = 0.17$, $\left<\log_{10}(M_1/M_2)\right> = 0.18$ and $\sigma(\log_{10}(M_1/M_2)) = 0.16$, giving $T_1 = 0.97$ and $T_2 = 0.87$. Recall that $T_1 \geq T_{1, \textrm{lim}} = 1$ and $T_2 \geq T_{2, \textrm{lim}} = 0.82$ would suggest that BCGs are statistical extremes rather than objects with a special mass (or luminosity) distribution. The values that we have found do not allow us to conclusively determine the statistical nature of BCGs in our sample. We repeat this test with our sample divided into redshift bins of width $\Delta z = 0.1$ to search for redshift evolution of the statistical nature of the BCGs. The results are shown in Figure~\ref{fig:t_stats_z_evol}. We find that the $T_1$ statistic generally indicates that the BCGs in our sample are special, while the $T_2$ statistic generally suggests the opposite, i.e. that they are statistical extremes. 

\begin{table*}
\caption{Tremaine-Richstone statistics for the full cluster sample and a number of subsamples based on the median cluster redshift, stellar mass and richness of the full sample.}
\centering
\begin{tabular}{|l|rrrrr|}
\hline
{\bf Sample} & \multicolumn{1}{c|}{$\sigma(\log_{10} M_1)$} & \multicolumn{1}{c}{$\sigma(\log_{10}(M_1/M_2))$} & \multicolumn{1}{c}{$\left<\log_{10}(M_1/M_2)\right>$} & \multicolumn{1}{c}{$T_1$} & \multicolumn{1}{c}{$T_2$} \\
\hline
Full sample & 0.174 & 0.157 & 0.180 & 0.968 & 0.870 \\
$z<0.656$ & 0.172 & 0.158 & 0.182 & 0.944 & 0.865 \\
$z\geq0.656$ & 0.175 & 0.156 & 0.178 & 0.985 & 0.875 \\
$\log_{10}(M_{\mathrm{tot}})<12.41$ & 0.162 & 0.158 & 0.180 & 0.899 & 0.878 \\
$\log_{10}(M_{\mathrm{tot}})\geq12.41$ & 0.149 & 0.155 & 0.180 & 0.826 & 0.862 \\
$N_{\mathrm{mem}}<19.33$ & 0.176 & 0.160 & 0.184 & 0.954 & 0.868 \\
$N_{\mathrm{mem}}\geq19.33$ & 0.169 & 0.153 & 0.176 & 0.962 & 0.871 \\
\hline
\end{tabular} 
\label{tab:dndzfits}
\end{table*}

\begin{figure}
    \centering
    \includegraphics[width=\columnwidth]{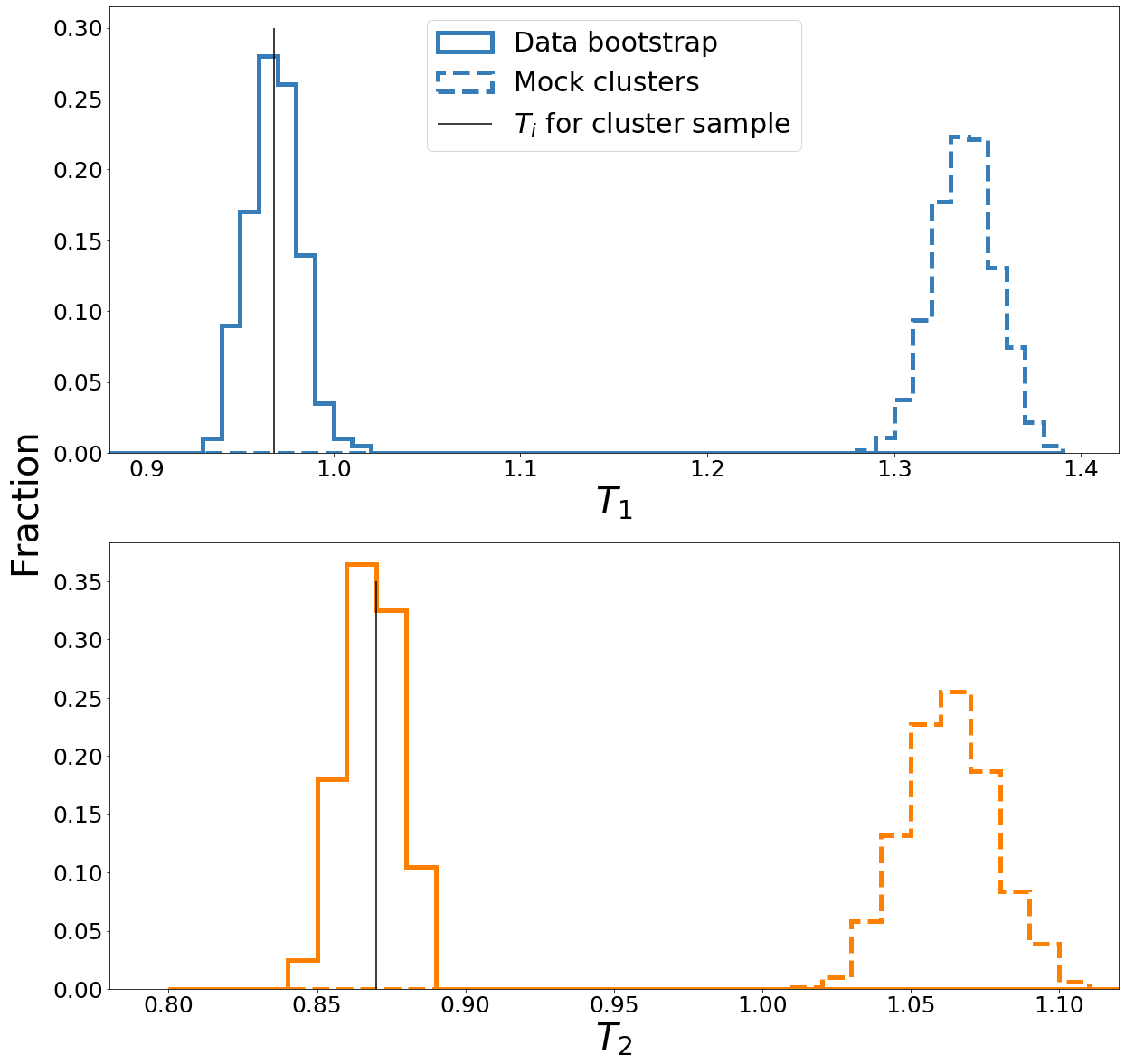}
    \caption{Distribution of the values of the Tremaine-Richstone statistics for 200 bootstrap resamplings of the clusters in our sample (solid histogram, with the true value indicated by a solid black line) and 1000 mock clusters ensembles (dashed histogram). There is no overlap between the two distributions; the $T_i$ values for the cluster sample are consistently below those for the mock clusters. This allows us to conclude that the masses of BCGs are special.}
    \label{fig:t_stat_dist_all}
\end{figure}

Noting that the limits $T_{1, \textrm{lim}} = 1$ and $T_{2, \textrm{lim}} = 0.82$ are dependent on the likely incorrect assumption of a power law form for the mass distribution, we adapt our analysis to be independent of the shape of the mass distribution. We construct ensembles of mock clusters as described in Section~\ref{sec:mass_corr_method}, select the two most massive galaxies in each mock cluster to be the BCG (with mass $M_1$) and the second-ranked galaxy (with mass $M_2$), and then compute the $T_1$ and $T_2$ statistics for each ensemble of mock clusters. The process by which these mock clusters are constructed explicitly assumes that their BCG has a mass that is drawn from the mass distribution of other cluster member galaxies. This allows us to determine whether the BCGs in our observed clusters are consistent with this assumption by comparing the $T_1$ and $T_2$ values from the mock clusters to those from our observed cluster sample. Specifically, we compare the distribution of $T_1$ and $T_2$ values from 1000 mock cluster realizations to the statistical uncertainty in the measurements of $T_1$ and $T_2$ from our observed clusters, derived from 200 bootstrap resamplings of the cluster sample. Figure~\ref{fig:t_stat_dist_all} shows the distribution of $T_1$ and $T_2$ values for 1000 mock cluster ensembles and 200 bootstrap resamplings of our true cluster sample. The distributions are converged with respect to the number of resamplings, i.e. they are not significantly different when using a larger number of resamplings. These distributions allow us to conclusively state that the Tremaine-Richstone statistics for our cluster sample are consistently below the values one would expect if BCGs were statistical extremes of the empirical mass distribution of cluster member galaxies. That is, the BCGs in our sample are unambiguously special. 

\begin{figure}
    \centering
    \includegraphics[width=\columnwidth]{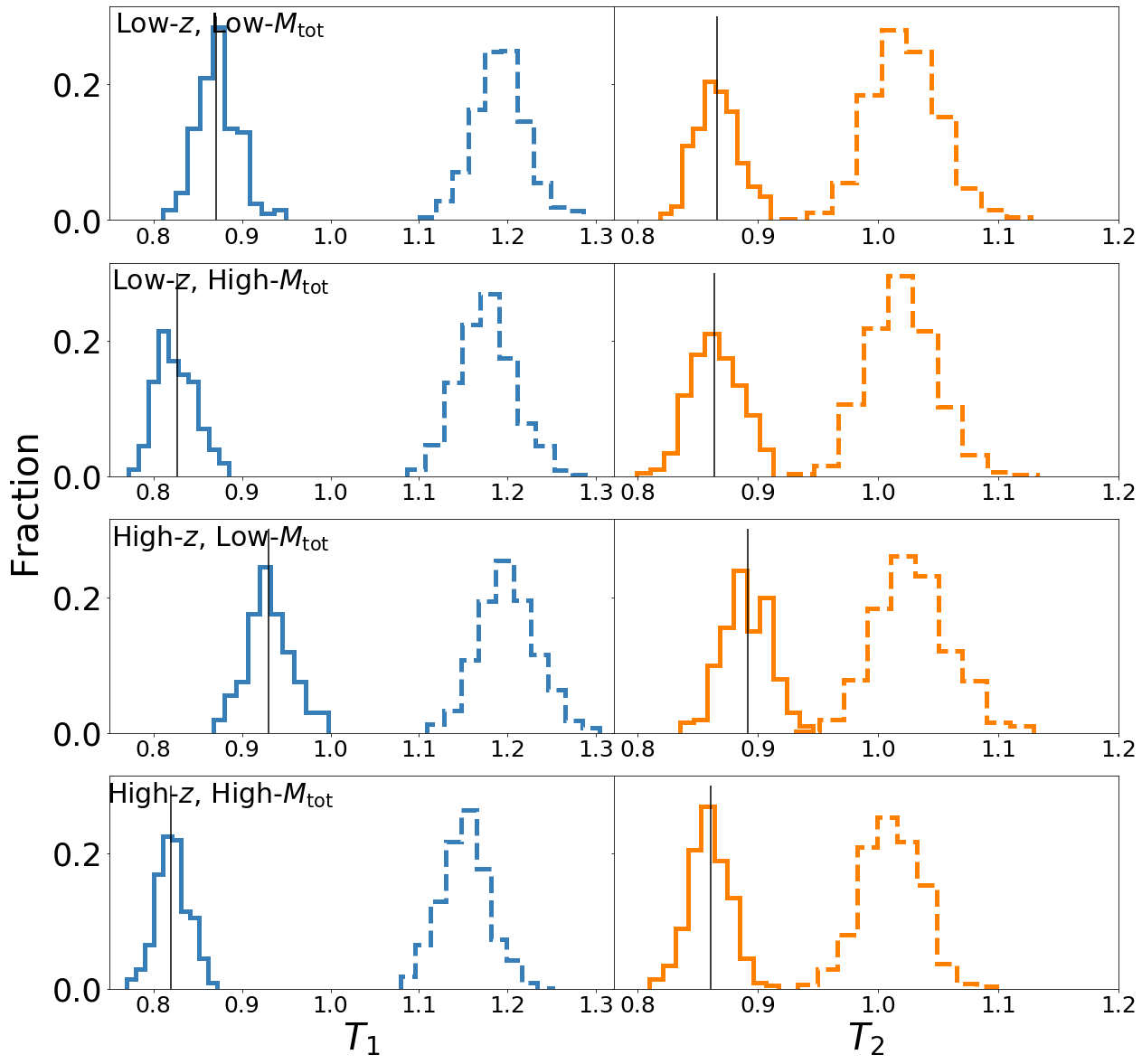}
    \caption{As in Figure~\ref{fig:t_stat_dist_all}, the distribution of the values of the Tremaine-Richstone statistics for various subsamples of the CAMIRA clusters (based on 200 bootstrap resamplings, solid histogram) and the corresponding mock cluster ensembles (from 1000 ensembles, dashed histogram). From top to bottom, the subsamples in each panel are low redshift and low cluster mass, low redshift and high cluster mass, high redshift and low cluster mass, and finally high redshift and high cluster mass. The subsamples are defined based on the median redshift and cluster stellar mass of our sample ($z=0.656$, $\log_{10} M_{\mathrm{tot}} = 12.41$). There is little to no overlap between the two distributions in each subsample, with the $T_i$ values for the true data being consistently below those for the mock clusters.}
    \label{fig:t_stat_dist_subsamples}
\end{figure}

We repeat this analysis dividing our clusters into four subsamples of high and low redshift and stellar mass, based on the median redshift and cluster stellar mass of our sample ($z=0.656$, $\log_{10} M_{\mathrm{tot}} = 12.41$). The galaxies used to populate the mock clusters in each of these cases are drawn from a pool of galaxies belonging to the subsample of clusters under consideration. For example, a galaxy from a high redshift cluster will not be used in a mock version of a low redshift cluster. This ensures that the true and mock $T_i$ statistics are derived from the same galaxy population, and the only difference is that the mock clusters are constructed assuming that the BCG mass is drawn from the cluster member mass distribution. These results are shown in Figure~\ref{fig:t_stat_dist_subsamples}. We come to the same conclusion for each of these subsamples, namely the distributions of both $T_1$ and $T_2$ are lower than what would be expected if BCGs were statistical extremes. Thus, we conclude that the BCGs in our sample are special at all redshifts and cluster stellar masses, with no obvious dependence on these quantities.  

\subsection{\texorpdfstring{$M_{\textrm{bcg}}-M_{\textrm{tot}}$} \ \ correlation}
\label{sec:mass_corr_results}

\begin{figure}
    \centering
    \includegraphics[width=\columnwidth]{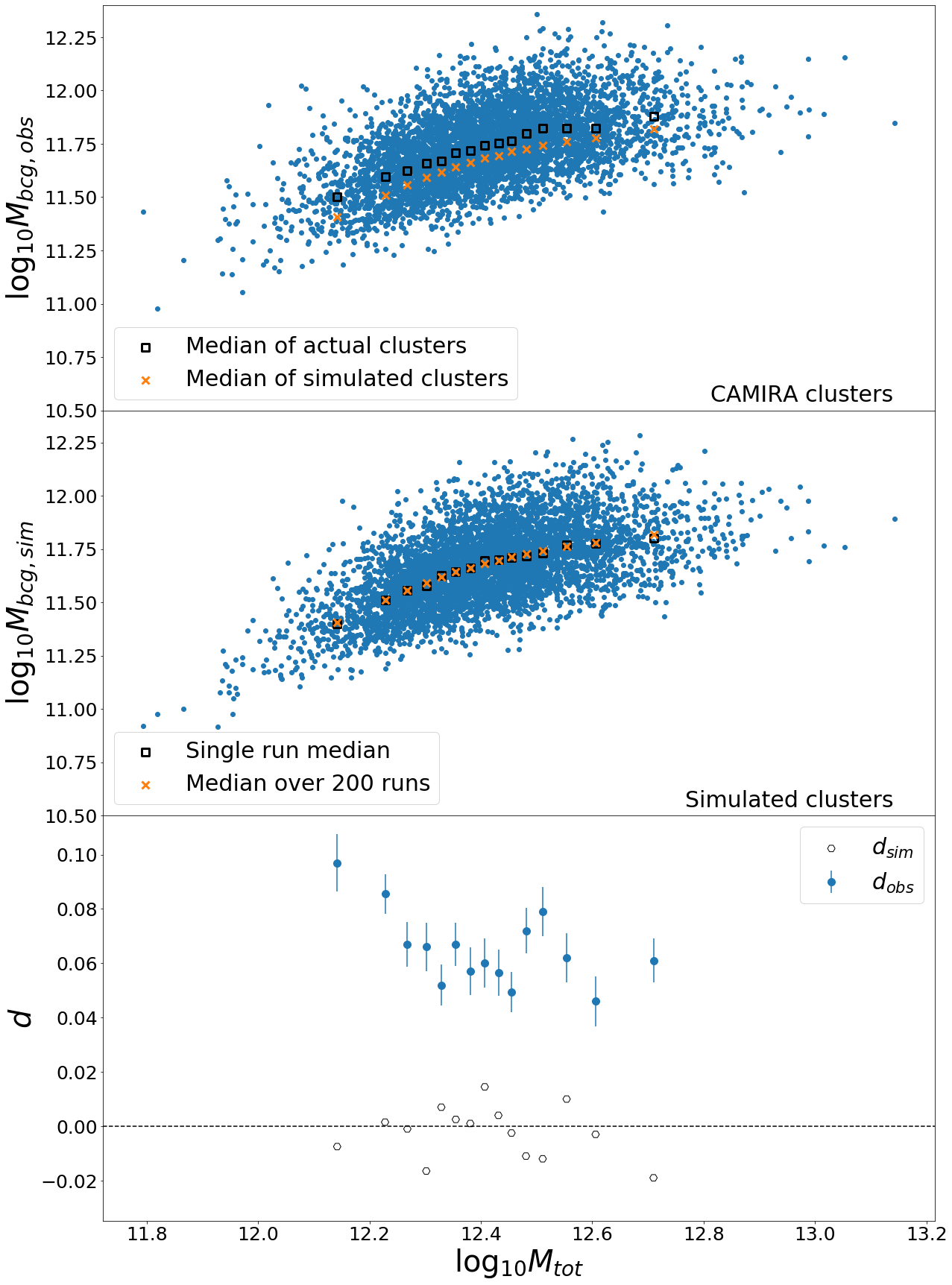}
    \caption{$M_{\textrm{bcg}}-M_{\textrm{tot}}$ correlation for the CAMIRA clusters (top panel) and one of the Monte Carlo realizations of the mock cluster ensemble (middle panel). The squares in the top two panels are the median of the BCG mass in equi-populated cluster mass bins, $\overline{M_{\textrm{bcg, obs}}}$ in the top panel and $\overline{M_{\textrm{bcg, sim}}}$ in the middle panel. The crosses in the top two panels are the median simulated BCG masses in each $M_{\textrm{tot}}$ bin, averaged over 200 Monte Carlo realizations, $\left<\overline{M_{\textrm{bcg, sim}}}\right>$. The bottom panel shows $d_{\textrm{obs}} \equiv \log_{10} \overline{M_{\textrm{bcg, obs}}} - \log_{10} \left<\overline{M_{\textrm{bcg, sim}}}\right>$ (blue solid points) and $d_{\textrm{sim}} \equiv \log_{10} \overline{M_{\textrm{bcg, sim}}} - \log_{10} \left<\overline{M_{\textrm{bcg, sim}}}\right>$ (black open points). The error bars on $d_{\textrm{obs}}$ come from 200 bootstrap resamplings of the observed $M_{\textrm{bcg}}-M_{\textrm{tot}}$ correlation. We find that BCGs are special independent of cluster stellar mass.}
    \label{fig:lin_2010_all_z}
\end{figure}

Figure~\ref{fig:lin_2010_all_z} shows the results from the observed $M_{\textrm{bcg}}-M_{\textrm{tot}}$ correlation in our sample compared to that from the simulated clusters (as described in Section~\ref{sec:mass_corr_method}), In creating the simulated cluster sample, we assumed that BCG masses are drawn from the same mass distribution as other cluster member galaxies. We use 200 realizations of the mock cluster ensemble, constructed as described in Section~\ref{sec:mass_corr_method} to calculate the median BCG mass in equi-populated bins of cluster mass $\overline{M_{\textrm{bcg, sim}}}$ for each realization and then the median of this over all realizations, $\left<\overline{M_{\textrm{bcg, sim}}}\right>$. We define the difference between the two as $d_{\textrm{sim}} \equiv \log_{10} \overline{M_{\textrm{bcg, sim}}} - \log_{10} \left<\overline{M_{\textrm{bcg, sim}}}\right>$, and we find that $d_{\textrm{sim}}$ roughly scatters about zero, as expected. We can also find the median of the BCG mass in each cluster mass bin for the CAMIRA clusters, $\overline{M_{\textrm{bcg, obs}}}$ and compute $d_{\textrm{obs}} \equiv \log_{10} \overline{M_{\textrm{bcg, obs}}} - \log_{10} \left<\overline{M_{\textrm{bcg, sim}}}\right>$. The error bars on $d_{obs}$ are derived from 200 bootstrap resamplings of the $M_{\textrm{bcg}}-M_{\textrm{tot}}$ correlation for true clusters in our sample. If BCG masses were drawn from the same mass distribution as other galaxies in the cluster, we would expect $d_{\textrm{obs}}$ to also be scattered about zero. However, we see that the values are consistently larger than zero in all cluster mass bins, indicating that the BCGs in our sample are more massive than those in the mock clusters, and are therefore more massive than one would expect if they are simply statistical extremes of the cluster member mass distribution. A Kolmogorov-Smirnov (KS) test shows that the probability of $d_{\textrm{obs}}$ being drawn from the same distribution as $d_{\textrm{sim}}$ is effectively 0, with a p-value of $10^{-8}$. 

\begin{figure}
    \centering
    \includegraphics[width=\columnwidth]{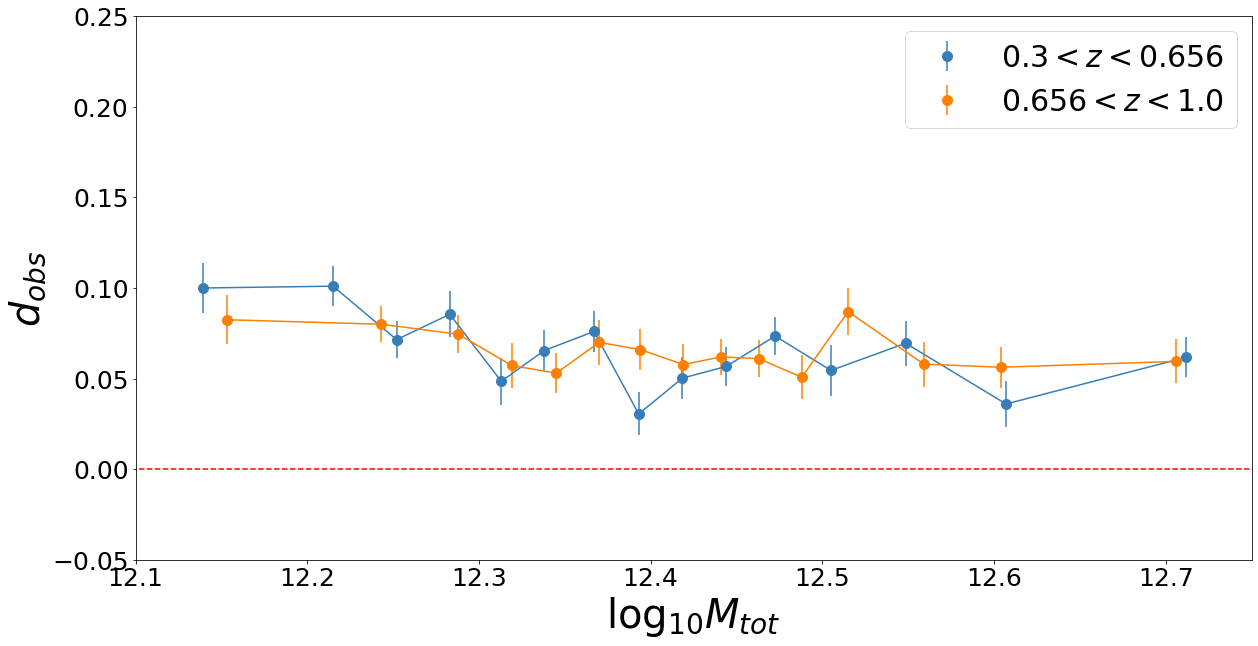}
    \caption{Value of $d_{\textrm{obs}} \equiv \log_{10} \overline{M_{\textrm{bcg, obs}}} - \log_{10} \left<\overline{M_{\textrm{bcg, sim}}}\right>$ in equi-populated bins of cluster mass split at the median redshift of the sample ($z=0.656$). We do not see any evidence for redshift evolution of the special nature of the BCGs in our sample.}
    \label{fig:lin_2010_low_vs_high_z}
\end{figure}

We repeat this analysis in bins of redshift to understand whether there is any redshift evolution in the statistical nature of the BCG mass. As described in Section~\ref{sec:t_stats_results}, the mock clusters are populated by galaxies in the same redshift bin as the CAMIRA clusters under consideration. Figure~\ref{fig:lin_2010_low_vs_high_z} shows $d_{\mathrm{obs}}$ as a function of cluster mass for low and high redshift subsamples based on the median redshift ($z=0.656$) of our full cluster catalog. We find no evidence of the deviation of the BCG mass from the expected mass changing with redshift. We also repeated this analysis with redshift bins of width $\Delta z =0.1$ and once again found no evidence for the evolution of the $d_{\mathrm{obs}}$ statistic with redshift. Thus, the $d_{\textrm{obs}}$ statistic also allows us to conclude that the BCGs in our sample are special independent of redshift and cluster stellar mass.

\section{Systematic Tests}
\label{sec:systematics}

We explore the robustness of our results upon relaxing and tightening various assumptions we have made to construct our cluster sample, including the minimum galaxy weight required to be considered a member of the cluster and the maximum blendedness fraction for the BCGs. We also split our cluster sample based on the BCG proximity to the cluster center to understand the effect of different environments and potential mergers on the special nature of BCGs. Finally, we explore the sensitivity of our results to projection effects. 

\subsection{Cluster membership probabilities}
\label{sec:weights}

\begin{figure}
    \centering
    \includegraphics[width=\columnwidth]{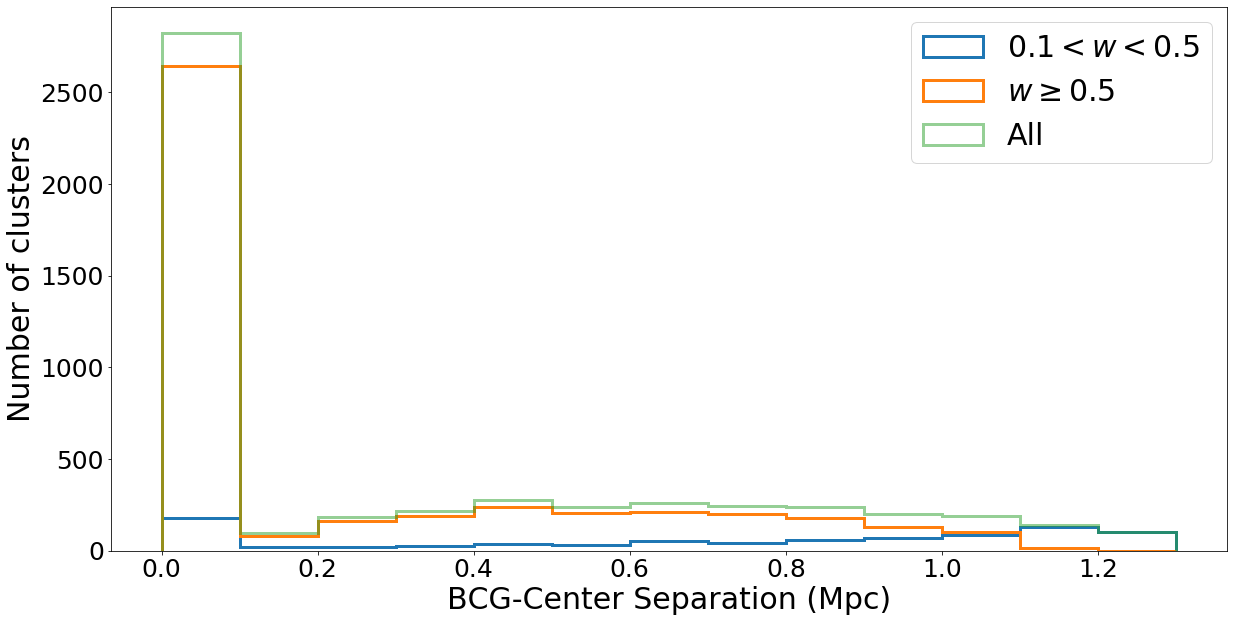}
    \caption{Distribution of projected distance of the BCG from the CAMIRA cluster center for BCGs in our sample with weights $w\geq0.5$ and those with weights $0.1<w<0.5$. Lower weighted BCGs are more likely to be further from the cluster center than than those with higher weights.} 
    \label{fig:offset_weight_dist}
\end{figure}

In the analysis above, we include all galaxies with a weight $w>0.1$ as cluster members. The weight for each galaxy is computed as described in Section~\ref{sec:cluster_gal_data}. It is the likelihood of each galaxy being a red sequence galaxy at redshift $z$ convolved with a mass filter (which is very close to 1 for all the galaxies in our sample) and a spatial filter. The spatial filter downweights galaxies with a larger projected distance from the cluster center, where the cluster center is defined to be the position of a massive galaxy near the center of the peak in the richness map. The spatial filter is, in practice, what the weights of galaxies in our sample are most sensitive to. Using the lower weight requirement of $w>0.1$ allows for massive galaxies further from the cluster center to be chosen as BCGs. This is illustrated in Figure~\ref{fig:offset_weight_dist}, which shows the distribution of offsets for BCGs with $w\geq0.5$ and those with $0.1<w<0.5$. 

\begin{figure}
    \centering
    \includegraphics[width=\columnwidth]{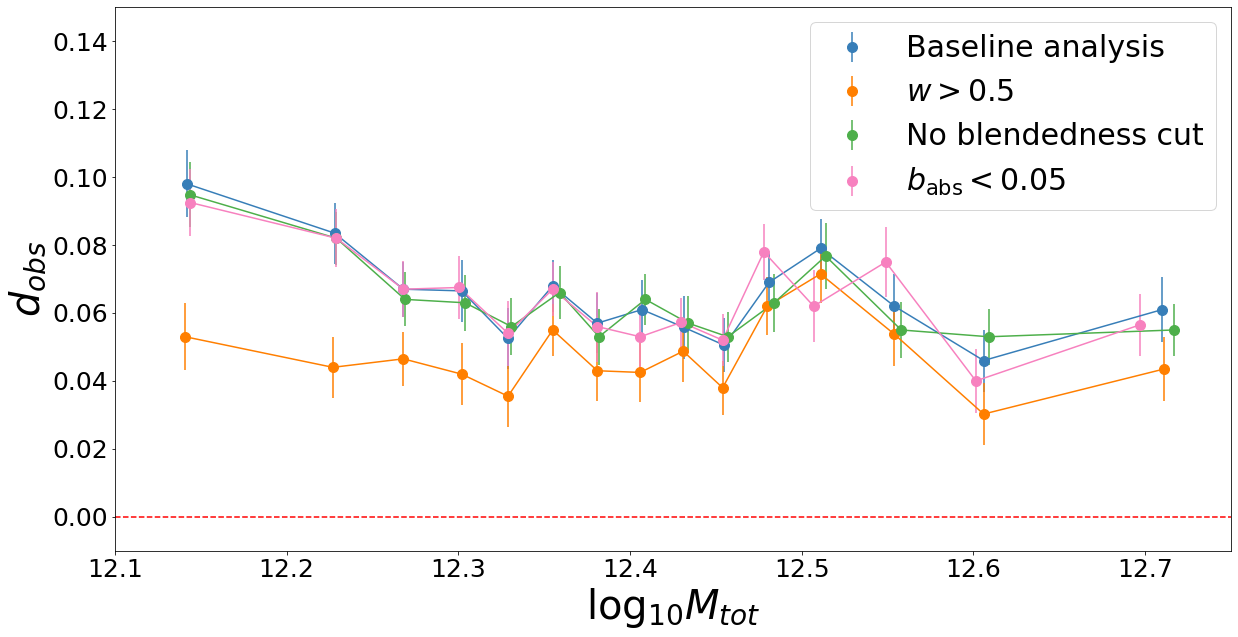}
    \caption{$d_{\textrm{obs}} \equiv \log_{10} \overline{M_{\textrm{bcg, obs}}} - \log_{10} \left<\overline{M_{\textrm{bcg, sim}}}\right>$ as a function of cluster mass for our baseline analysis, with $w>0.1$ and $b_{\textrm{abs}} < 0.1$ (blue), for a minimum galaxy weight of $w>0.5$ (orange), for $b_{\textrm{abs}} < 0.05$ (pink) and no blendedness cut (green). Qualitatively, our results remain the same across these different samples. BCGs are consistently more massive than would be expected if they were drawn from the cluster member mass distribution.} 
    \label{fig:lin_2010_systematics}
\end{figure}

We repeat the analyses of Section~\ref{sec:results} with a more stringent minimum weight of $w>0.5$. This causes the identification of $\sim 17\%$ of the BCGs in our sample to change. For the full cluster sample, we find $T_1 = 0.95$ and $T_2 = 0.84$, values that are slightly lower than those obtained when using $w>0.1$. When comparing the distribution of $T_1$ and $T_2$ from bootstrap realizations of the data to Monte Carlo simulations of clusters, we find no qualitiative difference in the results from the analysis above. There is still no overlap between these distributions, indicating that BCGs are not drawn from the same mass distribution as other cluster galaxies. Figure~\ref{fig:lin_2010_systematics} shows the results of the $M_{\textrm{bcg}}-M_{\textrm{tot}}$ correlation test for the BCG sample with $w>0.5$ in comparison with the results from our baseline analysis with $w>0.1$ (shown in Figure~\ref{fig:lin_2010_all_z}). Again, the qualitative results are the same; BCGs are consistently more massive than would be expected if they were drawn from the cluster member mass distribution. However, $d_{\textrm{obs}}$ is smaller when a more stringent weight cut is employed, particularly at low cluster masses. We explore possible reasons for this, in the context of the separation of the BCG from the cluster center, in Section~\ref{sec:offset}. 

\subsection{Photometry}
\label{sec:blendedness}

The results presented in Section~\ref{sec:results} are based on a sample that does not include BCGs with a blendedness fraction $b_{\textrm{abs}}>0.1$, as we would expect the estimated magnitudes, and hence the inferred masses of these objects to be more sensitive to errors in the deblending process. We explore the effect of changing this blendedness cut by making it more stringent (requiring $b_{\textrm{abs}}<0.05$) as well as by removing it altogether, i.e. including all BCGs in the analysis, regardless of blendedness fraction. In the former case, we find $T_1 = 0.98$ and $T_2 = 0.88$ and for the latter,  $T_1 = 0.97$ and $T_2 = 0.87$. These are very close to the values found when we required $b_{\textrm{abs}}<0.1$, and when we repeat the test of comparing the distributions of $T_1$ and $T_2$ from the data to those of simulated clusters, we once again find no overlap, indicating that the BCGs are not drawn from the cluster member mass distribution. The results from the $M_{\textrm{bcg}}-M_{\textrm{tot}}$ correlation test are shown in Figure~\ref{fig:lin_2010_systematics}. The values of $d_{\textrm{obs}}$ are very similar across these three samples with different blendedness cuts, and we find that BCGs are consistently more massive than would be expected if they were drawn from the cluster member mass distribution.

In addition to blending, we also consider the effect of uncertainties in the photometry and photometric redshift determination. \citet{oguri_2017} compare the estimated cluster photometric redshifts to spectroscopic redshifts for BCG identified by CAMIRA, using overlaps between HSC and a number of spectroscopic surveys (see Section 2.2 of \citet{oguri_2017} for details). The resulting scatter in the photometric redshift errors is $\sigma((z_{\mathrm{ph}}-z_{\mathrm{sp}})/(1+z_{\mathrm{sp}})) \sim 0.01$, which would translate to a $\sim 2\%$ uncertainty in the galaxy masses. Furthermore, the fraction of clusters that have a difference between the spectroscopic and photometric redshift of more than $5\sigma$ is only 0.49\%. We do not expect photometric redshift uncertainites to be the dominant source of error in the the inferred stellar masses. 

The formal quoted errors in the \textit{i} band \texttt{cmodel} magnitudes of the BCGs in our sample are typically a few percent or less, and are dominated by photon statistics. However, the \texttt{cmodel} magnitude does not necessarily capture the total magnitude of a galaxy as the local sky estimation can over-subtract the background, leading to an underestimation of the galaxy stellar mass. This might be of particular concern in galaxies like BCGs that often have extended envelopes. To determine whether such underestimation occurs and whether it has any impact on our results, we compare the inferred masses of the BCGs and other cluster member galaxies in our sample to the catalog by \citet{huang2018} (hereafter H18), which characterizes the light profiles of a sample of $\sim 7000$ massive galaxies with $0.3<z<0.5$ in the HSC data out to 100 kpc. The light profiles are determined via elliptical isophote fitting, and the resulting luminosity is converted to a stellar mass via SED fitting (see H18 for details). We expect the masses determined via this method to be a more accurate estimate of the total stellar mass of these galaxies, as the diffuse light at the outskirts of these galaxies is better accounted for. We cross-match both our BCG and cluster member galaxy catalog, where the latter includes the second-ranked galaxy, with the H18 catalog. Upon doing so, we find that the 100 kpc aperture stellar mass is, on average, about 0.1 dex higher than the CAMIRA stellar mass. However, this difference is approximately constant across the range of masses covered by the H18 catalog, i.e. the difference is similar for BCGs, second-ranked galaxies and other massive member galaxies (the H18 catalog is complete down to $\log(M_{*, \mathrm{cModel}}/M_{\odot}) \sim 11.5$). Since the statistics employed in our analysis, both $T_i$ and $d_{\textrm{obs}}$, depend only on the \textit{difference} of the log of galaxy masses, a systematic bias in the log of the estimated stellar mass that is consistent across mass will not affect our results. Although we are only able to do this comparison for objects in the redshift range $0.3<z<0.5$, we would expect the \texttt{cmodel} bias to become smaller at higher redshift, as the angular size of BCGs becomes smaller and cosmological dimming will make the extended halo quite faint. We conclude that uncertainties in the photometry and photometric redshifts do not have a significant impact on our results.

\subsection{BCG offset from cluster center}
\label{sec:offset}

\begin{figure}
    \centering
    \includegraphics[width=\columnwidth]{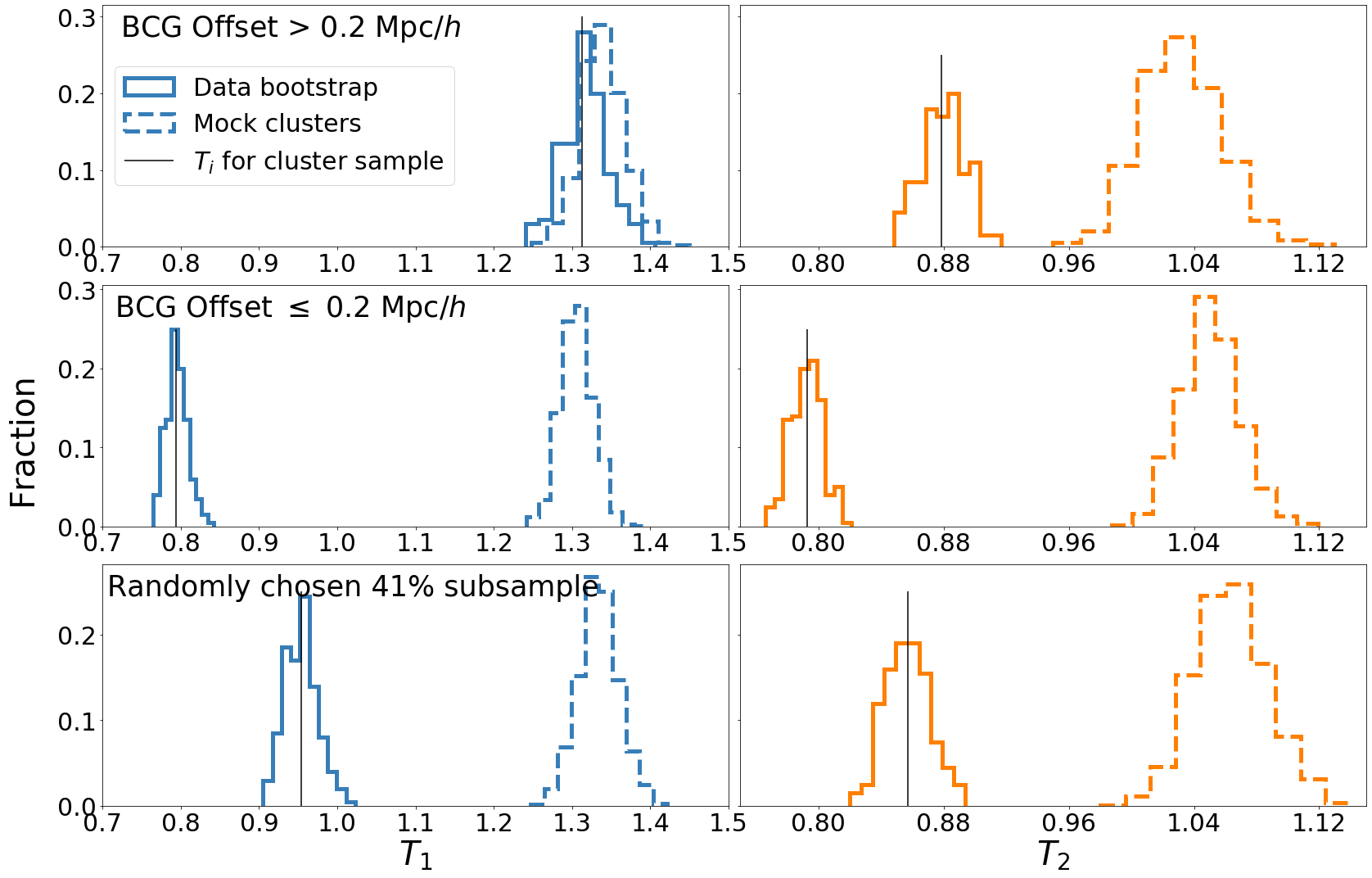}
    \caption{Distribution of the values of the Tremaine-Richstone statistics for the clusters in the $w>0.1$ sample (based on 200 bootstrap resamplings, solid histogram) and the mock clusters (from 1000 ensembles, dashed histogram). The true value for each subsample is indicated by a solid black line. The top panel is the subsample of clusters in which the BCG has a large offset from the cluster center, while the middle panel is the remaining clusters in the CAMIRA sample, i.e. the small offset sample. The bottom panel shows the results for a randomly drawn subsample of the cluster data, chosen to match the number of clusters in the large offset subsample. The small offset sample shows no overlap in $T_i$ values for the true data and the mock clusters, i.e. the Tremaine-Richstone statistics consistently indicate that the BCGs in small offset clusters are special. However, the large offset sample appears to be special based on the $T_2$ statistic, but is consistent with being a statistical extreme of the cluster member mass distribution based on the $T_1$ statistic.}
    \label{fig:t_stat_dist_offset}
\end{figure}

\begin{figure}
    \centering
    \includegraphics[width=\columnwidth]{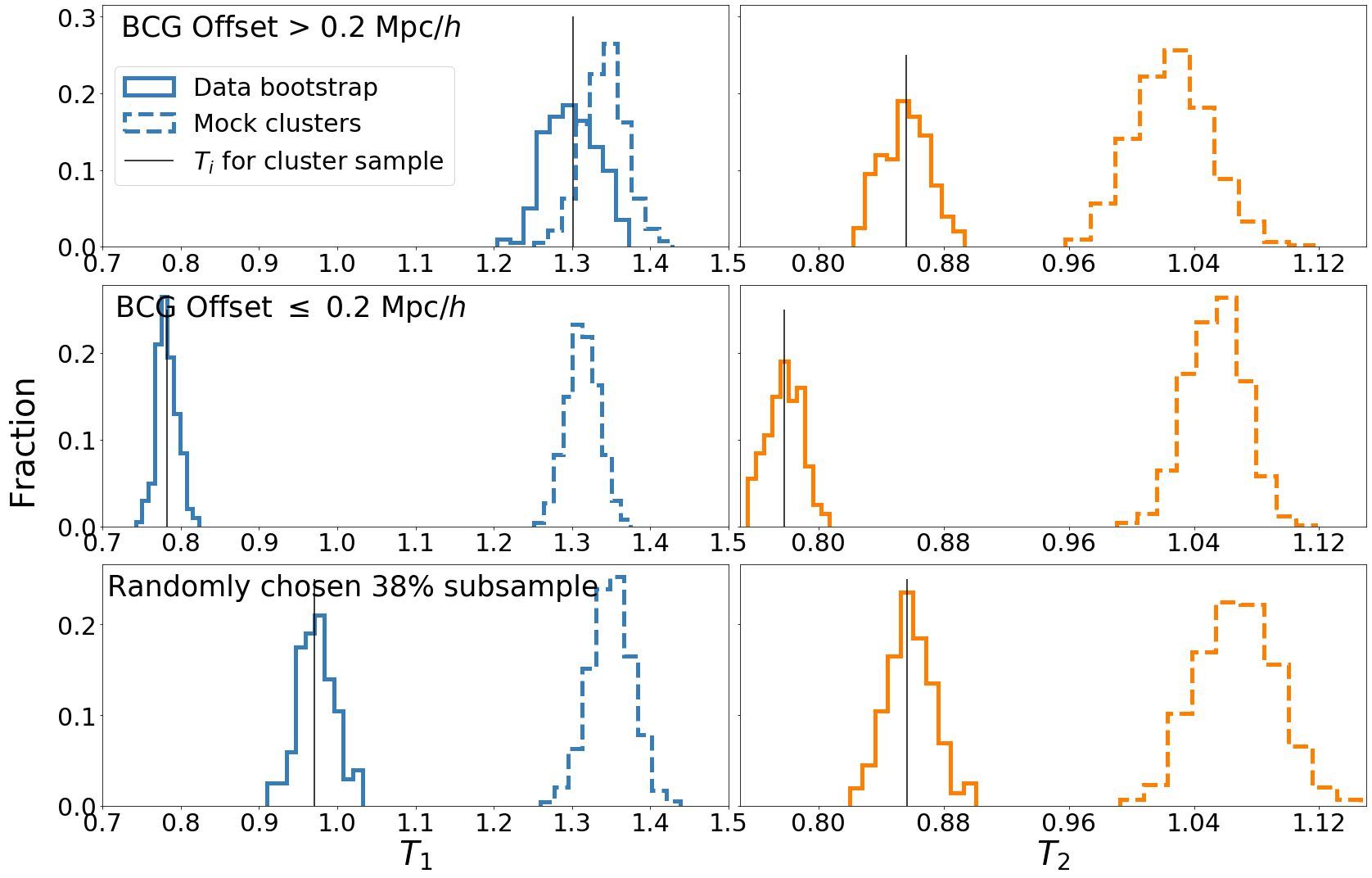}
    \caption{Distribution of the values of the Tremaine-Richstone statistics for the clusters in the $w>0.5$ sample. These results are consistent with those shown in Figure~\ref{fig:t_stat_dist_offset} for a minimum weight $w>0.1$. The small offset BCGs are inferred to be special from both $T_1$ and $T_2$, while for the large offset BCGs, $T_2$ suggests that they are special, whereas $T_1$ indicates that they are consistent with being a statistical extreme of the cluster member mass distribution.}
    \label{fig:t_stat_dist_offset_w_0.5}
\end{figure}

\begin{figure}
    \centering
    \includegraphics[width=\columnwidth]{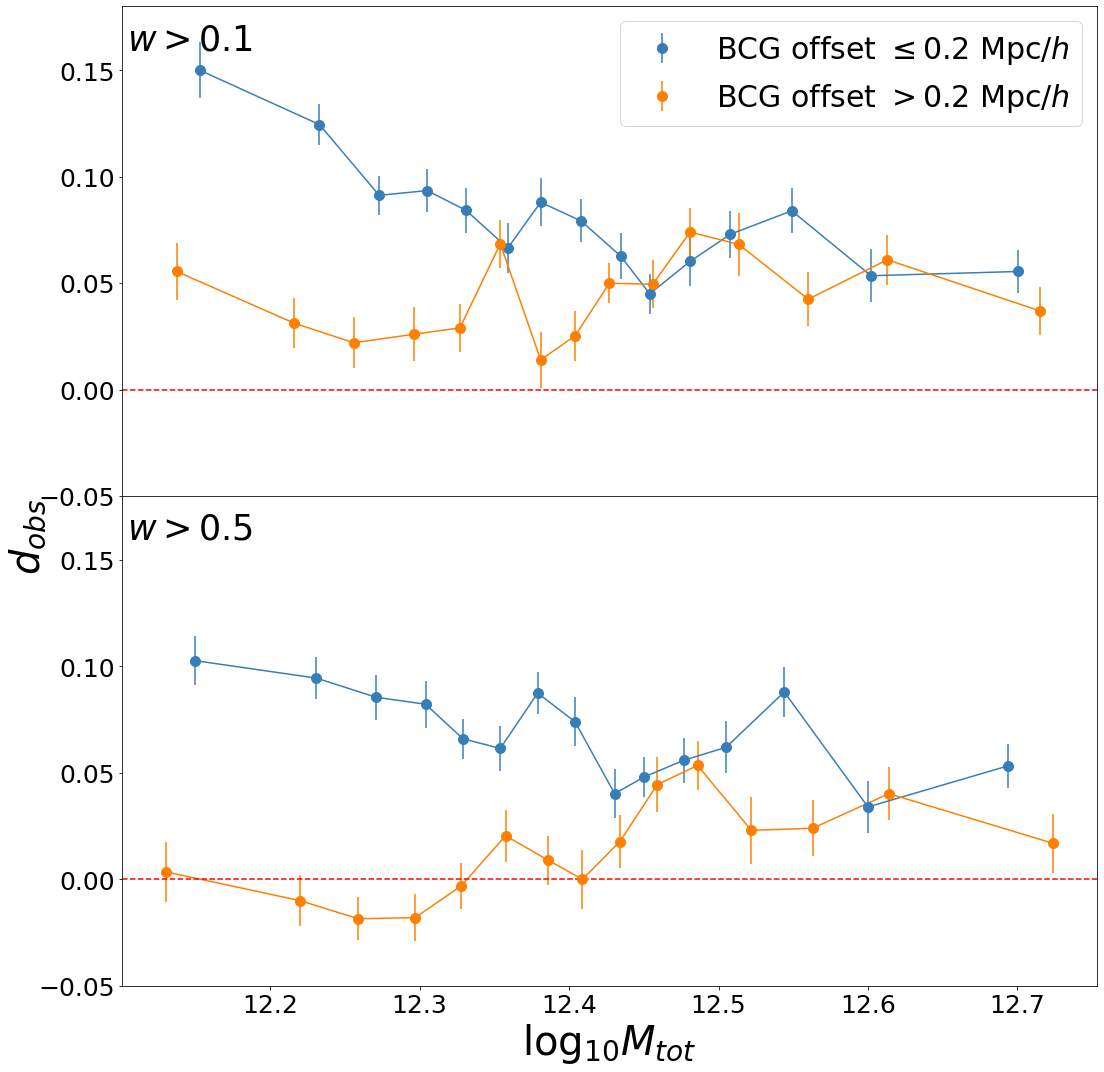}
    \caption{Value of $d_{\textrm{obs}}$ as a function of cluster mass for two subsamples of the CAMIRA clusters defined based on the projected separation of the BCG from the cluster center. The top panel shows the results from requiring a minimum weight of $w>0.1$ for cluster members. Large offset BCGs are seen to be consistent with being statistical extremes of the cluster member mass distribution at low cluster masses when using a more stringent weight cut. Small offset BCGs appear to always be special. }
    \label{fig:lin_2010_offset}
\end{figure}

If the luminosities and masses of BCGs are influenced by their environments, we would expect BCGs at the center of their clusters to be noticeably different from BCGs that are significantly offset from their cluster centers, since cluster centers are more densely populated. Furthermore, a BCG at the center of its cluster may indicate a more relaxed, virialized cluster, while BCGs with a large offset from the cluster center might be an indication of a cluster undergoing a merger. To test the effect of BCG position within the cluster, we divide the CAMIRA clusters into two subsamples, based on the projected separation of the BCG from the cluster center reported by CAMIRA. CAMIRA chooses the cluster center to be at the position of a massive galaxy (not necessarily the BCG that we have identified) near the richness peak. We define the large offset subsample to be the clusters for which the BCG-center separation is larger than $0.2$ Mpc $h^{-1}$, which is true for $\sim 41 \%$ of the clusters in our sample when using a minimum weight of $w>0.1$ and for $\sim 38 \%$ of the clusters when using $w>0.5$. The small offset subsample are those objects with a separation $\leq 0.2$ Mpc $h^{-1}$. The results of the Tremaine-Richstone tests for the large and small offset samples are shown in Figure~\ref{fig:t_stat_dist_offset} for the $w>0.1$ case and Figure~\ref{fig:t_stat_dist_offset_w_0.5} for the $w>0.5$ case. We include a third control sample in each of these tests, which represents a randomly selected fraction of the overall sample, where that fraction is similar to the large offset BCG fraction. This represents a random assignment of `large offset' or `small offset' to each cluster, which allows us to confirm that the different results that we see for the true large offset sample are robust. Figure~\ref{fig:lin_2010_offset} shows the results of the $M_{\textrm{bcg}}-M_{\textrm{tot}}$ correlation test for both the large and small offset subsamples. 

We see that regardless of the minimum weight requirement, the $T_1$ statistic suggests that BCGs with a large offset from the cluster center are statistical extremes, while those closer to the center are special. On the other hand, the $T_2$ statistic suggests that BCGs in both subsamples are special. For $w>0.1$, the KS statistic of the $T_1$ distribution for the large offset sample and the mock clusters is 0.33 with a p-value of $10^{-16}$. For $w>0.5$, the equivalent KS statistic is 0.51, with a p-value of $10^{-38}$. The KS statistic for the $T_2$ distributions of the large offset sample is 1.0 in both cases, with p-values of $\sim 10^{-145}$. 

Although we do not see significant changes based on the choice of minimum weight for the Tremaine-Richstone statistics, the $d_{\textrm{obs}}$ values are lower for the $w>0.5$ case than for $w>0.1$. For $w>0.1$, $d_{\textrm{obs}}$ is always greater than zero, indicating that BCGs in both the large and small offset subsamples are special, i.e. more massive than would be expected if they were drawn from the cluster member mass distribution. For $w>0.5$, we see that the mass of BCGs in the large offset sample at low cluster masses is consistent with being drawn from the mass distribution of other cluster galaxies. Small offset BCGs are found to be special at all cluster masses regardless of the minimum weight used. These BCGs are even more massive than expected, compared to the large offset subsample, particularly at low cluster masses. It is possible that the $w>0.1$ large offset sample includes contamination from massive foreground or background galaxies, which might explain why we see a signal in $d_{\textrm{obs}}$ for this sample, but not for the $w>0.5$ large offset sample. We explore the impact of projection effects in Section~\ref{sec:projection}. 

These results generally suggest that unlike BCGs that are near the cluster center, BCGs which are further away have masses that are more consistent with being drawn from the mass distribution of other cluster galaxies. This might be explained by BCGs in the small offset subsample growing in mass by cannibalizing small galaxies in their vicinity. From a galaxy-halo connection point of view, this means that a halo with a central BCG assembled earlier than one with the same cluster mass but an offset BCG, and enough time has elapsed since the last major halo merger that the cluster itself is relatively relaxed. The large offset clusters might have had a recent halo merger, or have ongoing halo assembly. Such a conclusion is supported by \citet{martel2014}, who used N-body simulations with a semi-analytical subgrid treatment of galaxy formation, merging, and tidal destruction to show that major and semi-major mergers of clusters are responsible for off-center BCG positions. The original BCGs of clusters that are merging will have similar stellar masses, leading to a small mass difference between the first and second-ranked galaxy, and hence a larger value of $T_1$. These results are also consistent with the findings of \citet{lauer2014}, who showed that the log-slope of the BCG photometric curve of growth, $\alpha$, tends to decrease with increased BCG offset from the cluster center, indicating that BCGs were puffed up by processes like mergers taking place in the inner region of clusters.

An interesting question to ask is whether a BCG with a large offset from the cluster center is indicative of a second-ranked galaxy near the center that might be the true BCG (with the more massive galaxy being an interloper), or the BCG of an original, un-merged cluster. We find that generally over 50\% of clusters with a BCG offset larger than $\sim 0.2$ Mpc $h^{-1}$ have a second-ranked galaxy that is also offset from the cluster center by more than $0.2$ Mpc $h^{-1}$. This suggests that many of these clusters might be undergoing a merger, such that the BCGs of the original clusters are both displaced from the center of the merged cluster.

However, large offset BCGs are also sensitive to other systematics in addition to projection effects, including mis-centering issues in the CAMIRA algorithm. Based on comparisons with clusters from the XMM Large Scale Structure survey \citep{pierre2004} and the XXL survey \citep{pierre2016}, the CAMIRA algorithm is found to have a mis-centering fraction of $\sim 32\%$ (based on both the S16A data \citep{oguri_2017} and the S20A data used in our analysis). That is, about one-third of the 31 CAMIRA clusters in our sample that can be cross-matched with the XXL catalog have an optically-identified cluster center that is significantly offset from the X-ray center. This was determined by fitting the distribution of offsets to a two-component Gaussian model. The `significantly offset’ clusters have a Gaussian distributed offset with a standard deviation of $0.26 \pm 0.04 h^{-1}$ Mpc, while the `well-centered' clusters have a standard deviation of $0.046 \pm 0.009 h^{-1}$ Mpc \citep{oguri_2017}. The mis-centering by the CAMIRA algorithm could  lead to an underestimation of the richness of the cluster, and therefore the cluster stellar mass values in Figure~\ref{fig:lin_2010_offset}, as well as incorrect determinations of the BCG offset from the cluster center.

We investigate the relationship between mis-centering and BCG position here, although we note we are limited by the small number of clusters in the CAMIRA sample that lie in the XXL field. About $58\%$ of the mis-centered clusters (based on the CAMIRA and X-ray centers) have BCGs with large offsets from the CAMIRA center. It is possible that a number of the BCGs from the large offset subsample are in fact closer to the X-ray center than the CAMIRA-identified cluster center, in which case they would still be at the bottom of the cluster potential well. However, this fraction is small; $\sim 20\%$ of the clusters whose CAMIRA centers are significantly ($> 0.2$ Mpc $h^{-1}$) different from the X-ray center have a BCG that is close to the X-ray center. These cases are contaminants in our large offset BCG sample, and could explain why $d_{\textrm{obs}}$ is not consistently zero over the range of cluster masses in Figure~\ref{fig:lin_2010_offset}. However, the mis-centering fraction when comparing the X-ray center to the BCG position is actually larger, about $39\%$, indicating that there is much that is still not understood about the mis-centering problem. 

\subsection{Projection effects}
\label{sec:projection}

Optical cluster finders like CAMIRA that rely on photometric data are susceptible to projection effects, where inaccurate photometric redshift estimates would lead the algorithm to mistakenly identify galaxies in distinct halos along the line of sight as a cluster, or erroneously assign galaxies along the line of sight to a bona-fide cluster. \citet{oguri_2017} use mock galaxy samples to evaluate the purity of the CAMIRA catalog, and find that $95\%$ of the CAMIRA-identified clusters in the mock catalogs have counterparts in the halo catalog, down to the richness limit of $N_{\textrm{mem}} = 15$. However, projection effects were not considered in their study. We assess the sensitivity of our results to contamination from projection effects by running a CAMIRA-like algorithm  on a galaxy catalog from the Millennium Simulation \citep{springel2005}. The algorithm is described in detail in \citet{murata2020}, and the galaxy catalog is described in \citet{sunayama2019}. 

The Millennium Simulation is an N-body simulation of a flat $\Lambda$CDM universe with $\Omega_m = 0.25$, $\Omega_b = 0.045$, $n_s = 1$, $\sigma_8 = 0.9$ and $H_0 = 73 \ \textrm{km} \ \textrm{s}^{-1} \textrm{Mpc}^{-1}$. The simulation uses $2160^3$ particles in a box of size $500 h^{-1}$ Mpc with a mass resolution of $8.6 \times 10^8 h^{-1} M_{\odot}$. Halos and their self-bound subhalos were identified using the \textit{subfind} algorithm \citep{springel2001}. The galaxy population was created using the semi-analytic model for galaxy formation in \citet{guo2011} and galaxies were selected from a snapshot at $z=0.24$ using the same threshold as in \citet{busch2017}. We require galaxies to have an \textit{i}-band absolute magnitude $M_i<-20.14$ and specific star formation rate below $1.5 \times 10^{-11} h \ \textrm{yr}^{-1}$. This gives us a sample of $\sim 925,000$ red galaxies with luminosity larger than $0.2L_{*}$. 

The primary difference between the true CAMIRA algorithm and the CAMIRA-like algorithm used here is in the color information present in the galaxies. Since the simulated galaxies do not have realistic color information, the algorithm employs a tophat filter along the line-of-sight around each halo center to mimic the photometric redshift uncertainty, with an effective projection length of $d_{\mathrm{eff}} = 40 h^{-1}$ Mpc. The projection length is chosen based on the typical standard deviation of the differences between the cluster redshift in CAMIRA and the spectroscopic BCG redshift (when available). This is roughly constant over $0.1 \leq z_{\mathrm{cl}} \leq 1.0$. Such an approach does not capture the effects of catastrophic photometric redshift errors, but we expect these to be negligible. Further details of the algorithm can be found in \citet{murata2020}. 

We create two sets of cluster catalogs from the Millennium Simulation. The first contains the true clusters, with the member galaxies being ones that are actually found inside the halos from the simulations, i.e within the virial radius. The second catalog is the observed cluster catalog, which contains the member galaxies and membership probabilities, as determined by the CAMIRA-like algorithm. We then apply both tests of the statistical nature of BCGs to each catalog, expecting to see the impact of projection effects in the results from the second catalog, but not those from the first. 

When computing the Tremaine-Richstone statistics, we find that $\sim 50\%$ of the second-ranked galaxies identified by the CAMIRA-like algorithm are interlopers, i.e. are not true members of the cluster. The BCGs identified in the CAMIRA-like catalog are always true members. This leads us to conclude that projection effects can lead to an overestimation of the second-ranked galaxy mass, and therefore an underestimation of the mass gap, i.e. the difference between the BCG and second-ranked galaxy masses. Since the Tremaine-Richstone statistics are dependent on the inverse of the mass gap, projection effects would lead to an overestimation of the values of these statistics. Indeed, the values of $T_1$ and $T_2$ estimated from the CAMIRA-like catalog ($T_1 = 0.79$, $T_2 = 0.74$) are larger than the values determined from the true cluster members from the Millennium simulations ($T_1 = 0.45$, $T_2 = 0.54$). Even if our sample is contaminated by projection effects, the estimated $T_1$ and $T_2$ statistics are already significantly below the values expected if BCG masses were consistent with being statistical extremes of the cluster member mass distribution. So, correcting for projection effects would only strengthen the significance of our results. We note that the estimate of contamination from projection effects ($\sim 50\%$ of misidentified second-ranked galaxies) derived using this method is very likely an overestimate. The uncertainty in photometric redshift correlates with the magnitude of galaxies, which is not accounted for with the simple tophat filter used in the CAMIRA-like algorithm. Thus, the 50\% fraction represents an upper limit on the degree of contamination.

\begin{figure}
    \centering
    \includegraphics[width=\columnwidth]{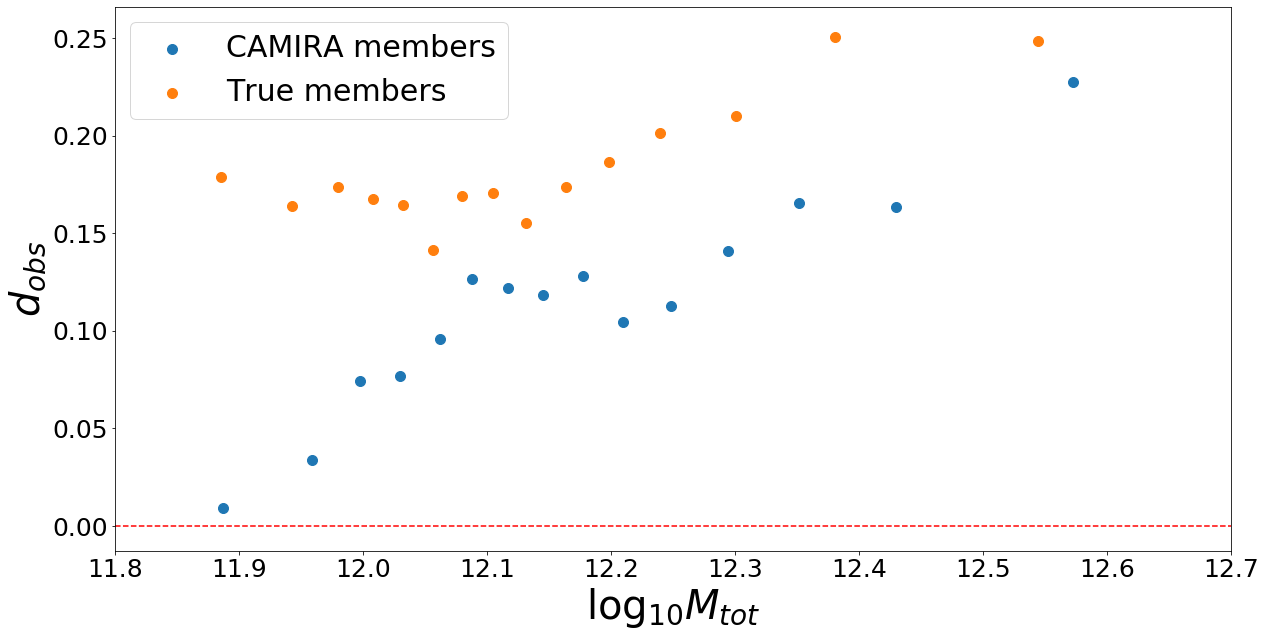}
    \caption{Value of $d_{\textrm{obs}}$ as a function of cluster mass for the true cluster member galaxies in the Millennium catalog, and the members identified by running the CAMIRA-like algorithm on the Millennium galaxy catalog. The catalog generated by the CAMIRA-like algorithm exhibits a cutoff in $d_{\textrm{obs}}$ at low cluster masses (around $10^{12.1} M_{\odot}$). This is not seen for the true members of the cluster, allowing us to conclude that it is the result of projection effects.}
    \label{fig:millennium_sims_dobs}
\end{figure}

We also apply our $M_{\textrm{bcg}}-M_{\textrm{tot}}$ correlation test to both the true member catalog and the CAMIRA-like catalog from the simulations. The resulting $d_{\textrm{obs}}$ for both catalogs is shown in Figure~\ref{fig:millennium_sims_dobs}. We find that the true cluster catalog from the simulations shows the same signal of BCG masses being consistently larger than what would be expected if they were drawn from the cluster member mass distribution. This suggests that the simulations include processes in their modeling of central galaxies that make them special at the redshift of the simulations, $z=0.24$ (we find that the central galaxy is the BCG in 99.86\% of the simulated clusters). We also find that the catalog generated by the CAMIRA-like algorithm exhibits a cutoff in $d_{\textrm{obs}}$ at low cluster masses, around $10^{12.1} M_{\odot}$, which is not seen in the true cluster catalog, nor in the true (observed) CAMIRA data, although the observed CAMIRA catalog does not reach such low cluster masses. Given that the CAMIRA-like algorithm identifies $\sim 4300$ clusters above the richness limit $N_{\textrm{mem}}>15$, while there are only $\sim 1800$ clusters in the Millennium catalog with more than 15 true member galaxies, we believe that projection effects cause the algorithm to mistakenly identify a a small group of galaxies as a cluster. In such low richness environments, we would not necessarily expect the ‘BCG’, or rather the brightest galaxy in this collection, to exhibit any special properties. This is more likely to happen at lower cluster stellar masses, as low mass clusters tend to have lower richness and are therefore more likely to be misclassified due to projection effects. This conclusion is supported by the larger comoving number density of the CAMIRA-like catalog, $3 \times 10^{-5} \ h^3 \ \textrm{Mpc}^{-3}$ compared to that of the true CAMIRA catalog which is less than $2 \times 10^{-5} \ h^3 \ \textrm{Mpc}^{-3}$ for the range of redshifts we consider here (see Figure 6 of \citealt{oguri_2017}). We do not see any cutoff in $d_{\textrm{obs}}$ at low cluster masses in our data. Furthermore, we find from the simulations that projection effects would likely dilute the signal of BCG specialness. The fact that we still see strong and consistent evidence for the special nature of BCGs across the range of cluster stellar masses in our data indicates that our results are likely not driven by projection effects. It is not unreasonable to conclude that the true CAMIRA algorithm is likely less susceptible to projection effects than the CAMIRA-like algorithm employed in this check, as the tophat filter is a crude approximation of the process by which CAMIRA uses color information to determine photometric redshifts. 

\section{Discussion and Conclusions}
\label{sec:discussion}

We have used a large sample of optically selected clusters to show via two different methods that the masses of BCGs are generally inconsistent with being drawn from the mass distribution of other cluster member galaxies. The Tremaine-Richstone statistics lead to inconsistent results when compared to the limiting values derived assuming a power-law form for the luminosity distribution. However, we simulate ensembles of mock galaxy clusters by assuming that the BCG is drawn from the mass distribution of other cluster member galaxies, and compare the resulting distributions of $T_1$ and $T_2$ to the values obtained from our data. In doing so, we are able to clearly show that the mass gap between the BCG and the second-ranked galaxy is too large to be consistent with BCGs being statistical extremes of the mass distribution of other cluster member galaxies. We also directly compare the BCG masses, as a function of cluster stellar mass, to the correlation obtained from the galaxy cluster simulations, and once again find that BCGs are consistently more massive than would be expected if they were drawn from the cluster member mass distribution. 

Upon dividing our sample into high- and low-redshift subsamples, we see no evidence for the evolution of the statistical nature of BCGs with redshift, i.e. BCGs appear to be drawn from a special mass distribution up to a redshift of at least $z=1.0$. Moreover, the numerical values of $T_1$, $T_2$ and $d_{\textrm{obs}}$  are essentially independent of redshift. This result is also robust as a function of cluster stellar mass.  This indicates that if BCGs were made `special' by a process like galactic cannibalism, it must have occurred earlier than $z=1.0$. We conclude that cannibalism significant enough to change the BCG demographics could not have happened at late times, and therefore models for BCG formation which include such late-time processes are not viable. If BCGs grew via major mergers, then these mergers must have taken place earlier than $z\approx1$. This is consistent with theoretical models of ``inside-out'' growth for BCGs, in which they grow via major mergers at early times and via minor mergers at late times \citep{DeLucia2007, vanDokkum2010, bai2014, Ragone-Figueroa2018}.

Future work will include extending this analysis to higher redshifts to understand whether there is an epoch at which galactic cannibalism or other processes took place to make BCGs special, or whether BCGs were born special. For example, \citet{sawicki2020} find that Ultra-Massive Passively Evolving Galaxies at $z\sim1.6$ are likely to be progenitors of the BCGs of present day galaxy clusters. They conclude that these progenitors have undergone major mergers at a redshift earlier than $z\sim1.6$ (as they appear to no longer have any massive companions), and will only continue to grow via minor mergers. Another data set that might be used to study high redshift BCGs is the sample from the Massive and Distant Clusters of WISE Survey (MaDCoWS), which includes 2433 clusters with $0.7 \lesssim z \lesssim 1.5$ in the Pan-STARRs field \citep{gonzelez2019}. However, the mass assembly of early-type galaxies in clusters is believed to take place at $z>1$, as little evolution is seen in their stellar mass functions up to $z \sim 1.2$ \citep{kodama2003, ellis2004, strazzullo2006}. Beyond a redshift of about $z\sim1.5$, very few virialized clusters are known. Instead, overdensities which are not yet virialized are often termed ``protoclusters''. \citet{kodama2007} study the color-magnitude relation of high redshift clusters and find that the bright end of the red sequence ($M_{\mathrm{stellar}} > 10^{11} M_{\odot}$) is well populated by $z \sim 2$, but is much less so in $z \sim 3$ protoclusters. This suggests that this analysis should be extended up to $z \sim 3$ to probe BCG specialness up to the time at which these galaxies were born. To do this, one could study high redshift radio galaxies (HzRGs), which are believed to be the progenitors of BCGs \citep{miley2008}. For example, \citet{wylezalek2013} identify a sample of $\sim 35$ clusters and protoclusters around radio-loud active galactic nuclei (AGN) between $z = 1.3$ and $3.2$. \citet{ito2019} identify 63 proto-BCGs at $z \sim 4$ from the brightest UV-selected galaxies in protoclusters, and find that these galaxies likely already had different star formation histories than other galaxies at the same epoch.

Upon dividing the clusters into large and small offset subsamples, based on the projected separation of the BCG from the cluster center (as identified by the CAMIRA algorithm), we find that BCGs with a large offset from the cluster center are more likely to be statistical extremes of the member galaxy mass distribution. This might be explained by the inability of a galaxy far from the center of the cluster potential to grow by galactic cannibalism, or by the large offset being indicative of merging clusters. In the case of a merger, the original BCGs from both clusters would remain, causing one of them to act as a "rival" BCG. We suggest that a definition of the BCG of a cluster that is based on its luminosity (or mass) as well as its proximity to the cluster center, rather than luminosity alone, may be more sensitive to the special statistical nature of this class of galaxies.

Additional data from from current and future surveys would help constrain BCG properties even further. The Vera C. Rubin Observatory \citep{ivezic2019} will provide a much larger optical cluster catalog, as it will cover a much larger solid angle than HSC, with $\sim 1$ mag greater depth. The next generation of spectroscopic surveys, including the Subaru Prime Focus Spectrograph \citep{takada14} and the Dark Energy Spectroscopic Instrument \citep{desi2016} will allow us to more accurately determine cluster membership, reducing potential impacts of projection effects and catastrophic photo-$z$ errors. Wide-field infrared imaging surveys, including Euclid \citep{amendola2013} and the Nancy Grace Roman Space Telescope \citep{spergel2015} will help push this analysis to higher redshifts, allowing us to probe the redshifts at which clusters begin to virialize. X-ray and Sunyaev-Zel'dovich cluster catalogs from surveys including eROSITA \citep{merloni2012} and the Atacama Cosmology Telescope \citep{hilton2021} will provide independent measurements of large numbers of cluster centers, allowing us to better study the mis-centering effect in CAMIRA, and thereby better understand the nature of BCGs that are significantly offset from the cluster center.

\section*{Acknowledgements}
We would like to thank Tod Lauer, Marc Postman, Marcin Sawicki and 
Lalitwadee Kawinwanichakij for helpful discussions and suggestions. We also thank the anonymous referee for helpful comments that improved the quality of this work.

The Hyper Suprime-Cam (HSC) collaboration includes the astronomical communities of Japan and Taiwan, and Princeton University. The HSC instrumentation and software were developed by the National Astronomical Observatory of Japan (NAOJ), the Kavli Institute for the Physics and Mathematics of the Universe (Kavli IPMU), the University of Tokyo, the High Energy Accelerator Research Organization (KEK), the Academia Sinica Institute for Astronomy and Astrophysics in Taiwan (ASIAA), and Princeton University. Funding was contributed by the FIRST program from Japanese Cabinet Office, the Ministry of Education, Culture, Sports, Science and Technology (MEXT), the Japan Society for the Promotion of Science (JSPS), Japan Science and Technology Agency (JST), the Toray Science Foundation, NAOJ, Kavli IPMU, KEK, ASIAA, and Princeton University.

This paper makes use of software developed for the Large Synoptic Survey Telescope. We thank the LSST Project for making their code available as free software at http://dm.lsst.org.

The Pan-STARRS1 Surveys (PS1) and the PS1 public science archive have been made possible through contributions by the Institute for Astronomy, the University of Hawaii, the PanSTARRS Project Office, the Max-Planck Society and its participating institutes, the Max Planck Institute for Astronomy, Heidelberg and the Max Planck Institute for Extraterrestrial Physics, Garching, The Johns Hopkins University, Durham University, the University of Edinburgh, the Queen’s University Belfast, the Harvard-Smithsonian Center for Astrophysics, the Las Cumbres Observatory Global Telescope Network Incorporated, the National Central University of Taiwan, the Space Telescope Science Institute, the National Aeronautics and Space Administration under Grant No. NNX08AR22G issued through the Planetary Science Division of the NASA Science Mission Directorate, the National Science Foundation Grant No. AST-1238877, the University of Maryland, E\"otv\"os Lor\'and University (ELTE), the Los Alamos National Laboratory, and the Gordon and Betty Moore Foundation.

This paper is based on data collected at the Subaru Telescope and retrieved from the HSC data archive system, which is operated by Subaru Telescope and Astronomy Data Center at National Astronomical Observatory of Japan. Data analysis was in part carried out with the cooperation of Center for Computational Astrophysics, National Astronomical Observatory of Japan.

RD acknowledges support from the NSF Graduate Research Fellowship Program under Grant No.\ DGE-2039656. Any opinions, findings, and conclusions or recommendations expressed in this material are those of the authors and do not necessarily reflect the views of the National Science Foundation. YTL is grateful for support from the Ministry of Science \& Technology of Taiwan under grant MOST 109-2112-M-001-005 and a Career Development Award from Academia Sinica (AS-CDA-106-M01).

\section*{Data Availability}

The data underlying this paper were accessed from the HSC collaboration. The derived data
generated in this research will be shared on a reasonable request to the first author.


\bibliographystyle{mnras}
\bibliography{bibliography} 







\bsp	
\label{lastpage}
\end{document}